\documentclass{svmult}

\usepackage{booktabs,url}
\usepackage{amsmath,amssymb}
\usepackage{graphicx}
\usepackage{mathptmx}       % selects Times Roman as basic font
\usepackage{helvet}         % selects Helvetica as sans-serif font
\usepackage{courier}        % selects Courier as typewriter font
\usepackage{makeidx}         % allows index generation
\usepackage{multicol}        % used for the two-column index
\usepackage[bottom]{footmisc}% places footnotes at page bottom

\DeclareMathOperator{\std}{std}

\begin{document}

% \citestyle{nature}

\title*{Modularity maximization and tree clustering: Novel ways to determine effective geographic borders}
\titlerunning{Novel ways to determine effective geographic borders}
\author{D. Grady, R. Brune, C. Thiemann, F. Theis, and D. Brockmann}
\institute{D. Grady \at Northwestern University, Evanston, IL
\and R. Brune \at Max Planck Institute for Dynamics and Self-Organization, Göttingen, Germany
\and C. Thiemann \at Max Planck Institute for Dynamics and Self-Organization, Göttingen, Germany
\and F. Theis \at Institute of Bioinformatics and Systems Biology, Helmholtz Zentrum München, German Research Center for Environmental Health, Neuherberg \& Department of Mathematical Science, Technische Universität München, Garching, Germany
\and D. Brockmann \at Northwestern University, Evanston, IL,  \email{brockmann@northwestern.edu}}
\maketitle

\abstract{Territorial subdivisions and geographic borders are essential for understanding phenomena in sociology, political science, history, and economics. They influence the interregional flow of information and cross-border trade and affect the diffusion of innovation and technology. However, most existing administrative borders were determined by a variety of historic and political circumstances along with some degree of arbitrariness. Societies have changed drastically, and it is doubtful that currently existing borders reflect the most logical divisions. Fortunately, at this point in history we are in a position to actually measure some aspects of the geographic structure of society through human mobility. Large-scale transportation systems such as trains and airlines provide data about the number of people traveling between geographic locations, and many promising human mobility proxies are being discovered, such as cell phones, bank notes, and various online social networks. In this chapter we apply two optimization techniques to a human mobility proxy (bank note circulation) to investigate the effective geographic borders that emerge from a direct analysis of human mobility.}
\abstract*{Territorial subdivisions and geographic borders are essential for understanding phenomena in sociology, political science, history, and economics. They influence the interregional flow of information and cross-border trade and affect the diffusion of innovation and technology. However, most existing administrative borders were determined by a variety of historic and political circumstances along with some degree of arbitrariness. Societies have changed drastically, and it is doubtful that currently existing borders reflect the most logical divisions. Fortunately, at this point in history we are in a position to actually measure some aspects of the geographic structure of society through human mobility. Large-scale transportation systems such as trains and airlines provide data about the number of people traveling between geographic locations, and many promising human mobility proxies are being discovered, such as cell phones, bank notes, and various online social networks. In this chapter we apply two optimization techniques to a human mobility proxy (bank note circulation) to investigate the effective geographic borders that emerge from a direct analysis of human mobility.}

\section{Introduction}

The geographic compartmentalization of maps into coherent territorial units is not only essential for the management and distribution of administrative responsibilities and the allocation of public resources. Territorial subdivisions also serve as an important frame of reference for understanding a variety of phenomena related to human activity. Existing borders frequently correlate with cultural and linguistic boundaries or topographical features~\cite{Newman:2006p1735, Eaton:2002p1738}, they represent essential factors in trade and technology transfer~\cite{Ernst:2002p1742, Keller:2004p1752}, and they indirectly shape the evolution of human-mediated dynamic processes such as the spread of emergent infectious diseases~\cite{Ferguson:2005p1755, Colizza:2006p1756, Colizza:2007p1757, Lloyd20041}.

The majority of existing administrative and political borders, for example in the United States and Europe, evolved over centuries and typically stabilized many decades ago, during a time when human interactions and mobility were predominantly local and the conceptual separation of spatially extended human populations into a hierarchy of geographically coherent subdivisions was meaningful and plausible.

However, modern human communication and mobility has undergone massive structural changes in the past few decades~\cite{Newman:2006p1735, Lazer:2009p2180}. Efficient communication networks, large-scale and widespread social networks, and more affordable long-distance travel generated highly complex connectivity patterns among individuals in large-scale human populations~\cite{Balcan:2009p2707, Helbing:2007}. Although geographic proximity still dominates human activities, increasing interactions over long distances~\cite{Brockmann:2006p6, gonzalez2008nature, Song:2010p4432} and across cultural and political borders amplify the small-world effect~\cite{Watts:1998p9, LibenNowell:2005p606} and decrease the relative importance of local interactions.

\begin{figure}
    \centering
    \includegraphics[width=4in]% 
        {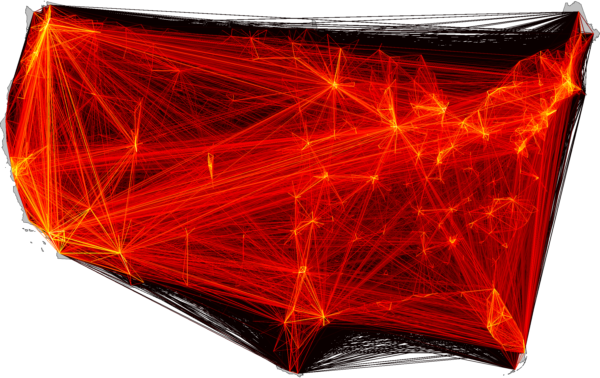}
    \caption{The Where's George? network. Multi-scale human mobility is characterized by dominant short-range and significant long-range connectivity patterns. The illustrated network represents a proxy for human mobility, the flux of bank notes between 3,109 counties in the lower 48 United States. Each link is represented by a line, the color scale encodes the strength of a connection from small (dark red) to large (bright yellow) values of $w_{ij}$ spanning four orders of magnitude.}
    \label{fig:WG}
\end{figure}

\begin{figure}
    \centering
    \includegraphics[width=0.9\textwidth]% 
        {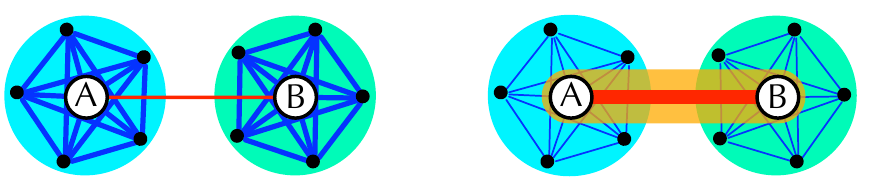}
    \caption{A simplified illustration of generic traffic patterns between and within metropolitan mobility hubs (A and B), with two types of connections $w_L$ and $w_D$, local traffic connecting individual hubs to smaller nodes in their local environment (blue) and long-distance links connecting the hubs (red). Depending on the ratio of local and long-range flux magnitude, two qualitatively different modularizations are plausible. If $w_L \gg w_D$, two spatially compact communities are meaningful (left), whereas if $w_L \ll w_D $, the metropolitan centers belong to one geographically delocalized module (orange), effectively detached from their local environment, yielding three communities altogether (right).}
    \label{fig:traffic_cartoon}
\end{figure}

Human mobility networks epitomize the complexity of multi-scale connectivity in human populations (see Figure~\ref{fig:WG}). More than 17~million passengers travel each week across long distances on the United States air transportation network alone. However, including all means of transportation, 80\% of all traffic occurs across distances less than 50\,km~\cite{Brockmann:2006p6, Brockmann:2008p1762}. The coexistence of dominant short-range and significant long-range interactions handicaps efforts to define and assess the location and structure of effective borders that are implicitly encoded in human activities across distance. The paradigm of spatially coherent communities may no longer be plausible, and it is unclear what structures emerge from the interplay of interactions and activities across spatial scales~\cite{Brockmann:2006p6, Vespignani:2009p1760, Brockmann:2008p1762, gonzalez2008nature}. This difficulty is schematically illustrated in Figure~\ref{fig:traffic_cartoon}. Depending on the ratio of local versus long-range traffic, one of two structurally different divisions of subpopulations is plausible. If short-range traffic outweighs long-range traffic, local, spatially coherent subdivisions are meaningful. Conversely, if long-range traffic dominates, subdividing into a single, spatially de-coherent urban community and disconnected suburban modules is appropriate and effective geographic borders are difficult to define in this case.

Although previous studies identified community structures in long-range mobility networks based on topological connectivity~\cite{Guimera:2005p1776, SalesPardo:2007p1500}, this example illustrates that the traffic intensity resulting from the interplay of mobility on all spatial scales must be taken into account. Obtaining comprehensive, complete, and precise datasets on human mobility covering many spatial scales is a difficult task, and recent studies have followed a promising alternative strategy based on the analysis of proxies that permit indirect measurement of human mobility patterns~\cite{Brockmann:2006p6, gonzalez2008nature, Song:2010p4432, Lazer:2009p2180, Eagle:2010p4790, Eagle:2009p4610}.

We focus on one human mobility proxy, a dataset collected at the website \url{www.wheresgeorge.com}. This website hosts a bill-tracking game called Where's George? in which participants can tag an individual US banknote of any denomination by logging in to the website and entering the bill's serial number along with their location. Subsequent participants who receive the bill may do the same, thereby recording a part of the spatial trajectory the bill follows during its lifetime. We use this information to construct a network whose nodes represent counties in the continental United States and whose edges encode the number of bills exchanged between pairs of counties; details of this and a discussion of some statistics of the data are given in Section~\ref{sec:wg}.

Both of our analyses rest on the idea of finding community partitions of the network, that is, dividing all of the nodes into a set of mutually disjoint groups or communities. A community of nodes can be defined in many different ways, but all definitions try to capture some aspect of the intuitive idea of a community: a set of nodes that belong together, or are more similar to one another than they are to the rest of the population.

Our first analysis in Section~\ref{sec:modmax} uses a modularity maximization technique to identify community partitions. Modularity is a method of scoring any given community partition in a network. A partition with a high modularity score has many more intra-group links, and fewer inter-group links, than expected by random chance. Our optimization algorithm searches for high-modularity partitions through a stochastic, simulated annealing process.

We go on to determine community partitions in Section~\ref{sec:shortest_path_tree_clustering} by searching for nodes with similar topological features, namely their shortest-path tree. Each node is the root of a shortest-path tree that comprises a minimal set of the strongest links connecting that node to the rest of the network. By looking for topological similarities between shortest-path trees, we identify groups of nodes that have similar patterns of connectivity.

With both methods, once a community partition is identified, a corresponding geographic border structure is produced simply by drawing borders between counties that do not belong to the same community, and in Section~\ref{sec:borders} we discuss how a superposition of border structures alleviates some of the long-standing weaknesses of modularity maximization. The fact that communities tend to be spatially compact is one of the most surprising findings of this research, and we conclude in Section~\ref{sec:comparisons} by developing a method for comparing border structures and examining the degree to which effective mobility borders line up with various existing borders, such as state boundary lines, census areas, and economic areas.

\section{Network modularity}\label{sec:modmax}

This section introduces the modularity measure and describes the simulated annealing algorithm we use for finding maximal-modularity community partitions. 

We assume here that $W$ is a square, symmetric matrix that represents a symmetric, weighted network; the elements $w_{ij}$ are nonnegative and measure the strength of the connection between nodes $i$ and $j$. Based on the idea that two nodes $i$ and $j$ are effectively proximal if $w_{ij}$ is large, we search for a community partition of the nodes that has a high value of modularity\cite{Danon:2005p1196, Girvan:2002p1192, Newman:2004p1191}. This standard network-theoretic measure of community structure prefers partitions such that the intra-connectivity of the modules in the partition is high and inter-connectivity between them is low as compared to a random null model. Given a partition $P$ of the nodes into $k$ modules $M_n$, the modularity $Q(P)$ is defined as
\begin{equation}\label{Qdef}
    Q = \sum_n \Delta F_n
\end{equation}
in which $\Delta F_n = F_n - F_n^0$ is the difference between $F_n$, the fraction of total mobility within the module $M_n$, and the expected fraction $F_n^0$ of a random network with an identical weight distribution $p(w)$. $Q$ cannot exceed unity; high values indicate that a partition successfully groups nodes into modules, whereas random partitions yield $Q \approx 0$. Maximizing $Q$ in large networks is an NP-hard problem~\cite{Brandes:2008p2590}, but a variety of algorithms have been developed to systematically explore and sample the space of possible divisions in order to identify high-modularity partitions~\cite{Danon:2005p1196, Fortunato:2010p2643}.

\subsection{Finding optimal partitions}

As discussed in more detail in Section~\ref{sec:degeneracy}, our method relies on finding several different high-modularity partitions, which restricts the range of applicable algorithms. For example, the deterministic divisive algorithms described by Newman and Girvan~\cite{Newman:2004p1191} cannot find several different local maxima of the modularity function. In contrast, Monte Carlo algorithms return different partitions with probabilities that monotonically increase with the corresponding modularity values, one of which is the simulated annealing algorithm described by Guimer\`{a} and Amaral~\cite{guimera2005jstat}. Additionally, this algorithm has been found to perform the best in terms of correctly identifying modules in networks with artificial community structure in a survey by Danon~et~al.~\cite{Danon:2005p1196}, which lead us to choose this algorithm for our work.

The partition vector $P$ is initialized such that each of the $N$ nodes is in its own module, $P_i = i$. Alternatively, one could randomly assign each node to one of a few modules to form the initial partition.  We found, however, that in this case the algorithm will split these few large modules into a large number of very small modules before slowly merging them into the final result.  Since splits of large modules, involving a recursive simulated annealing run, are computationally very expensive, we avoid them by starting with a partition of single-node modules.

A small modification of the partition is then made (see below) to obtain a new partition $P'$ and its effect on the modularity value, $\Delta Q = Q(P') - Q(P)$.  If $\Delta Q > 0$, the new partition is better than the old one and we replace $P = P'$.  If $\Delta Q < 0$, the partition is only accepted with probability $p_T(\Delta Q) = \exp(\Delta Q/T)$, where $T$ is a ``temperature'' that controls the typical penalty on $Q$ we are willing to accept with the new partition $P'$.

This procedure is repeated a number of times, initially with a high $T = T_0$ accepting modifications with large negative impact on modularity and therefore allowing to sample multiple local maxima.  After $O(N^2)$ modifications, the temperature is lowered by a \emph{cooling factor} $c$.  When $T$ is small enough, worse partitions are not accepted anymore and the partition $P$ has ``annealed'' into a local maxima of the modularity landscape $Q(\cdot)$.

During each temperature step, we intersperse $f\,N^2$ local with $f\,N$ global modifications of the partitions, where $f$ is a tuning parameter.  A local modification is a switch of one node to another, randomly selected, module, while a global modification can be a merge of two or a split of one randomly selected module.  Finding a suitable split of a module that is not immediately rejected is done by recursively running a simplified version of the simulated annealing algorithm on it: the module in question is extracted and treated as an independent network, initially randomly partitioned into two modules. Only local modifications are allowed while annealing this bipartition into a local modularity maximum. Afterwards, the split module is replaced into the full network and evaluated against the modularity value of the full partition.

We observed that the global structure of the partition is found quickly by the algorithm and mostly only local modifications are accepted at low temperatures.  Since the split operations are computationally intensive, we therefore track the number of rejected split modifications in each temperature step and reduce the probability of future trials if that number is high.

To generate the large ensemble of partitions discussed in Section~\ref{sec:borders}, we used $T_0 = 2.5\cdot 10^{-4}$ as initial temperature, $c = 0.75$ as the cooling factor, and $f = 0.05$.  We abort the procedure and accept the partition as ``optimal'' if no better partition is found in three consecutive temperature steps.

The run time of this stochastic algorithm depends in a complex way on both the size \emph{and} the structure of the the input, and therefore the time complexity does not scale with a simple function of the input size. However, we found the algorithm to perform very well and in acceptable runtime (60--90 minutes on a 2.8 GHz processor for most runs) with the configuration given above, although these parameters are less conservative than those proposed by Guimer\`a~et~al\cite{guimera2005jstat}. The large ensemble of resulting partitions (Figure~\ref{fig:ensemble_mod_stats}) has a tight distribution of modularity values, indicating the algorithm tends to converge onto a stable maximum.

\section{A proxy for multiscale human mobility networks} \label{sec:wg}

% TODO The distributions of link weight and node flux in WG are not included
% These were originally in the last panel of Figure 1 in the actual paper; the distance distribution is reproduced in this section, but weight and flux dists are not. We may not need to include them; just wanted to make a note.

Here we construct a proxy network for human mobility from the geographic circulation of banknotes in the United States. Movement data was collected using the online bill-tracking game at \url{www.wheresgeorge.com}. Individuals participating in this game can mark individual bills and return them to circulation; other individuals who randomly receive bills can report this find online along with their current location (zip code). Our analysis is based on the intuitive notion that the coupling strength between two locations $i$ and $j$ increases with $w_{ij}^H$, the number of individuals that travel between a pair of locations per unit time, and furthermore that the flux of individuals in turn is proportional to the flux of bank notes, denoted by $w_{ij}$. Evidence for the validity of this assumption has been obtained previously~\cite{Brockmann:2006p6, Brockmann:2008p1762, gonzalez2008nature} and we provide further evidence below.

As of January 15th, 2010 a total of $187,925,059$ individual bills are being tracked at the website \url{www.wheresgeorge.com}. Approximately 11.24\% of those have had {}``hits'', i.e.~they were reported a second time at the site after initial entry. The current analysis is based on a set of $N_{0}=11,950,239$ bills that were reported at least a second time. For each bill $n$ we have a sequence of pairs of data $$B_{n}=\{Z_{n,i},T_{n,i}\}\quad i=0,\ldots,L_{n}\quad n=1,\ldots,N_{0}$$ of zip codes $Z_{n,i}$ and times $T_{n,i}$ at which the bill was reported. Each $B_{n}$ reflects a geographic trajectory of a bill with $L_{n}$ individual legs. In total, we have $14,612,391$ single legs in our database. Note that the majority ($81.78\%$) of trajectories are single-legged reflecting a reporting probability of $\approx 20\%$ during the lifetime of a bill. 

The set of $B_{n}$ represents the core dataset of our analysis. For each bill we have additional information:
\begin{enumerate}
	\item Denomination: \$1, \$2, \$5, \$10, \$20, \$50, or \$100. The fraction of each denomination is depicted in Table~\ref{tab:Denominations-in-the}.
	\item The Federal Reserve Bank code, A through L, corresponding to one of 12 of the United States Federal Reserve Banks that issued the bill. The fraction of bills as a function of FRB origin is provided in Table~\ref{tab:Absolute-number-and}.
\end{enumerate}

\begin{table}
  \centering
  \caption{\label{tab:Denominations-in-the}Denominations in the WG dataset}
  \begin{tabular}{cccccccc}
  \toprule
  Denomination & \$1 & \$2 & \$5 & \$10 & \$20 & \$50 & \$100\\
  \midrule
  Number of bills & 9,931,261 & 36,639  & 1,069,427 & 401,101 & 461,076 & 24,209 & 26,526\\
  Fraction {[}\%{]} & 83.11 & 0.31 & 8.95 & 3.36 & 3.86 & 0.20 & 0.22\\
  \bottomrule
  \end{tabular}
\end{table}

\begin{table}
	\centering
	\caption{\label{tab:Absolute-number-and}Absolute number and relative fraction of bills based on Federal 	Reserve Bank}
	\begin{tabular}{clrr}
		\toprule
		FRB Code & Location & Count & Fraction {[}\%{]}\\
		\midrule
		A & Boston & 799,537 & 6.69\\
		B & New York City & 1,325,942 & 11.10\\
		C & Philadelphia & 822,340 & 6.88\\
		D & Cleveland & 661,278 & 5.53\\
		E & Richmond & 948,516 & 7.94\\
		F & Atlanta & 1,565,732 & 13.10\\
		G & Chicago & 1,207,448 & 10.10\\
		H & St. Louis & 472,930 & 3.96\\
		I & Minneapolis & 360,194 & 3.01\\
		J & Kansas City & 713,393 & 5.97\\
		K & Dallas & 869,866 & 7.28\\
		L & San Francisco & 2,203,063 & 18.44\\
		\bottomrule
	\end{tabular}
\end{table}

We restrict the analysis to the lower 48 states and the District of Columbia (thus excluding Hawaii and Alaska) and consider only legs with origin and destination locations in these states, reducing the original dataset to $11,759,420$ bills (98.40\% of the original data) and $14,376,232$ trajectory legs (98.38\%).

The spatial resolution of the dataset is given by $41,106$ zip codes, with mean linear extent of $14$\,km. The mean linear extent of the lower 48 states is $2,842$\,km defining the bounds of the system. For each zip code $Z_{i}$ we use centroid information to associate with each report a longitude/latitude location $\mathbf{x}=(\Theta,\phi)$, such that each trajectory $n$ corresponds to a sequence of geographic locations $\mathbf{X}_{i}$ with $i=1,\ldots,L_{n}$:
\begin{equation}
	t_{n}:\quad\left\{ \mathbf{X}_{n,0},\Delta T_{n,1},\mathbf{X}_{n,1},\ldots,\Delta T_{n,L_n},\mathbf{X}_{n,L_n}\right\} \quad\text{with}\quad n=1,\ldots,N_{0}
	\label{eq:trajectories}
\end{equation}
where $\mathbf{X}_{n,0}$ is the initial entry location, and $\Delta T_{n,i}=T_{n,i}-T_{n,i-1}$ are inter-report times.

\subsection{Geographical distributions}

\begin{figure}
    \centering
    \parbox[c]{.7\linewidth}{%
        \includegraphics[width=\linewidth]{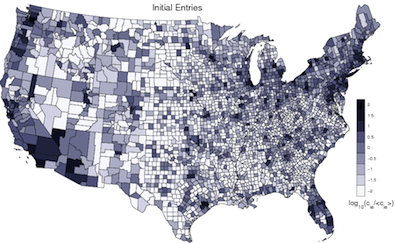}\\[\baselineskip]
        \includegraphics[width=\linewidth]{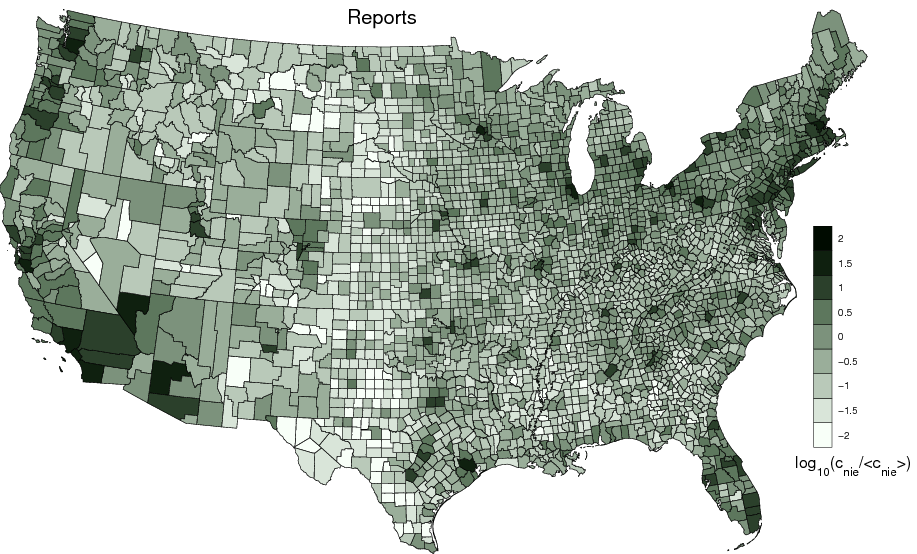}\\[\baselineskip]
        \includegraphics[width=\linewidth]{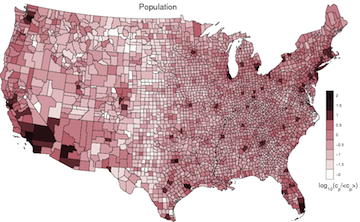}}
    \caption{\label{fig:grint}The frequencies of reports (top) and initial entries (middle) correlate with the county population (bottom) in the lower 48 states.}
\end{figure}

Based on these trajectories we define the density of initial entries as
\begin{equation}
	p_{IE}(\mathbf{x})=\frac{1}{N}\sum_{n=1}^{N}\delta(\mathbf{x}-\mathbf{X}_{n,0})
	\label{eq:initialentrydensity}
\end{equation}
and the density of reports as
\begin{equation}
	p_{R}(\mathbf{x})=\frac{1}{N}\sum_{n=1}^{N}\frac{1}{L_{n}}\sum_{i=1}^{L_{n}}\delta(\mathbf{x}-\mathbf{X}_{n,i}),
	\label{eq:reportdensity}
\end{equation}
where $\delta$ is the Dirac delta function, equal to $1$ when its argument is $0$ and equal to $0$ otherwise.

In order to assess the spatial distribution of reports and initial entries and to quantify the correlation with the population density we compute the number of reports and initial entries for each of the $M=3,109$ counties in the lower 48 states. Defining for each county $k$ a characteristic function
\begin{equation}
\chi_{k}(\mathbf{x})=\begin{cases}
1 & \quad\text{if}\quad\mathbf{x}\in P_{k}\\
0 & \quad\text{otherwise}\end{cases}
\label{eq:countyPoly}
\end{equation}
where $P_{k}$ is the polygon defining the county's interior, the number of reports and initial entries in county $k$ are given by
$$
	m_{R}(k) = \left\langle \chi_{k}\right\rangle _{R} = \int\chi_{k}(\mathbf{x})\,p_{R}(\mathbf{x})\,\text{d}\mathbf{x}\quad\text{and}\quad
	m_{IE}(k) = \left\langle \chi_{k}\right\rangle _{IE} = \int\chi_{k}(\mathbf{x})\,p_{IE}(\mathbf{x})\,\text{d}\mathbf{x},
$$
respectively. Figure~\ref{fig:grint} compares the distribution of reports $m_{R}(k)$, initial entries $m_{IE}(k)$ and the population $P(k)$ of the $3,109$ counties. As all three quantities are positive and vary over many orders of magnitude, the maps depict $\log_{10}(m_{R})$, $\log_{10}(m_{IE})$ and $\log_{10}(P)$. Qualitatively, reports and initial entries correlate strongly with the population density. Computing the correlation coefficient of the logarithmic quantities yields $c(R,P)=0.933$ and $c(IE,P)=0.819$. Despite the expected increase of $m_{R}(k)$ and $m_{IE}(k)$ with $P(k)$, only the report count increases approximately linearly with population size, whereas initial entries show a deviation for small populations. We believe that this deviation is a consequence of the social difference between the subpopulation of ``Georgers'' that are responsible for initiating bills and entering them into the system, ``actively'' playing the game, and the larger group of people that randomly receive a bill and report it, ``passively'' participating. This hypothesis could explain that areas with higher population densities contain a larger proportion of internet-savvy communities that are inclined to become Georgers and initiate bills. In order to exclude a potential bias caused by this effect we exclude all the legs in~(\ref{eq:noIEtrajectories}) that contain an initial entry as the origin, i.e.~we only investigate the reduced set
\begin{equation}
	t_{2,n}:\quad\left\{ \mathbf{X}_{n,1},\Delta T_{n,2},\mathbf{X}_{n,2},\ldots,\Delta T_{n,L_n},\mathbf{X}_{n,L_n}\right\} \quad\text{with}\quad n=1,\ldots,N_{0}
	\label{eq:noIEtrajectories}
\end{equation}
that excludes the first legs of all $t_{n}$. Excluding the first leg reduces the number of bills to $4,743,330$. However, the key results, e.g.~the border structures discussed in Section~\ref{sec:borders}, are robust against the inclusion of initial entries. Computing mobility networks based on either set, $t_{n}$ or $t_{2,n}$ does not change the observed pattern significantly.

\subsection{Distance and time: spatially averaged quantities}

\begin{figure}
	\sidecaption[c]
	\includegraphics[width=.5\columnwidth]{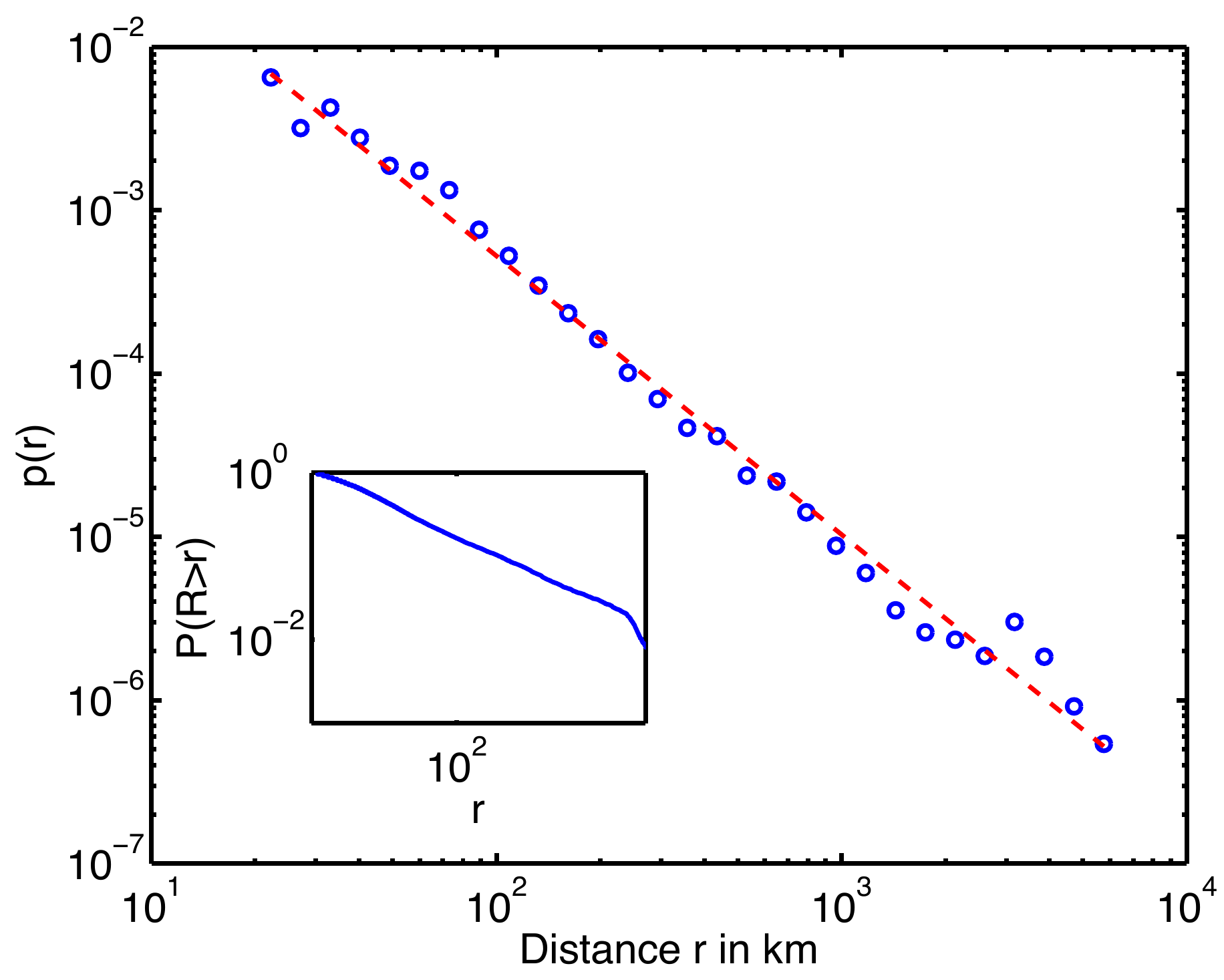}
	\caption{\label{fig:The-estimated-probabilityofr}The estimated probability $p(r|t<\tau)$ of a bill traversing a distance $r$ in time $t<\tau$ where $\tau=4$~days. In red a maximum likelihood fit of the the function $p(r)\sim r^{-(1+\beta)}$ with $\beta=0.7056$.}
\end{figure}

\noindent From $t_{2,n}$ we extract pairs of spatio-temporal leg distances $\{d_s(\mathbf{X}_{n,i},\mathbf{X}_{n,i-1}),\Delta T_{n,i}\}$, where $d_s(\cdot,\cdot)$ denotes the distance on a sphere (shorter segment of the great circle that passes through both points). This type of dataset was first investigated in 2006 based on a much smaller core dataset of bill trajectories~\cite{Brockmann:2006p6}. In particular, the combined probability density (pdf)
\begin{equation}
	p(r,t)=\left\langle \delta\left(r-d_s(\mathbf{X}_{n,i},\mathbf{X}_{n,i-1})\right)\delta\left(t-\Delta T_{n,i}\right)\right\rangle
	\label{eq:prt}
\end{equation}
was estimated as well as marginal pdfs $p(r)$ and $p(t).$ The central finding of the 2006 study was that $p(r)\sim r^{-(1+\beta)}$ and that the time evolution of the density~(\ref{eq:prt}) can be described by a bi-fractional diffusion equation. Here we reproduce some of the properties before we construct the mobility network used in the main text. Figure~\ref{fig:The-estimated-probabilityofr} shows the short time pdf of a bill traversing a distance $r$ in a time $t<\tau$ where we chose $\tau=4$~days. Using maximum likelihood we find this function can be described by a power-law \[p(r)\sim\frac{1}{r^{1+\beta}}\quad\text{with}\quad\beta=0.7056\pm0.0659.\] This power law describes the dispersal characteristics on a population-averaged level. The short-time distance pdf represents a dispersal kernel and for small times $t$ approximates the instantaneous rate of traversing a distance $r$. 

\begin{figure}
    \centering
    \includegraphics[width=0.49\columnwidth]{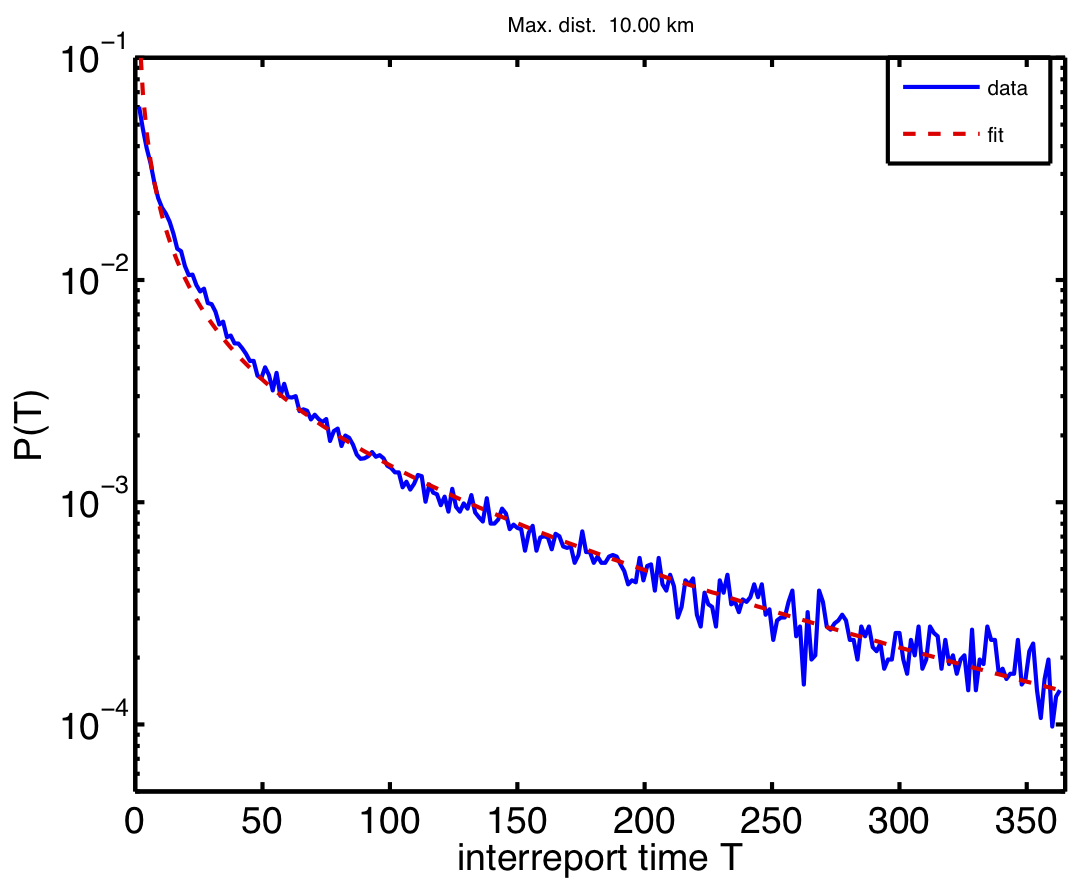}\hfill
    \includegraphics[width=0.49\columnwidth]{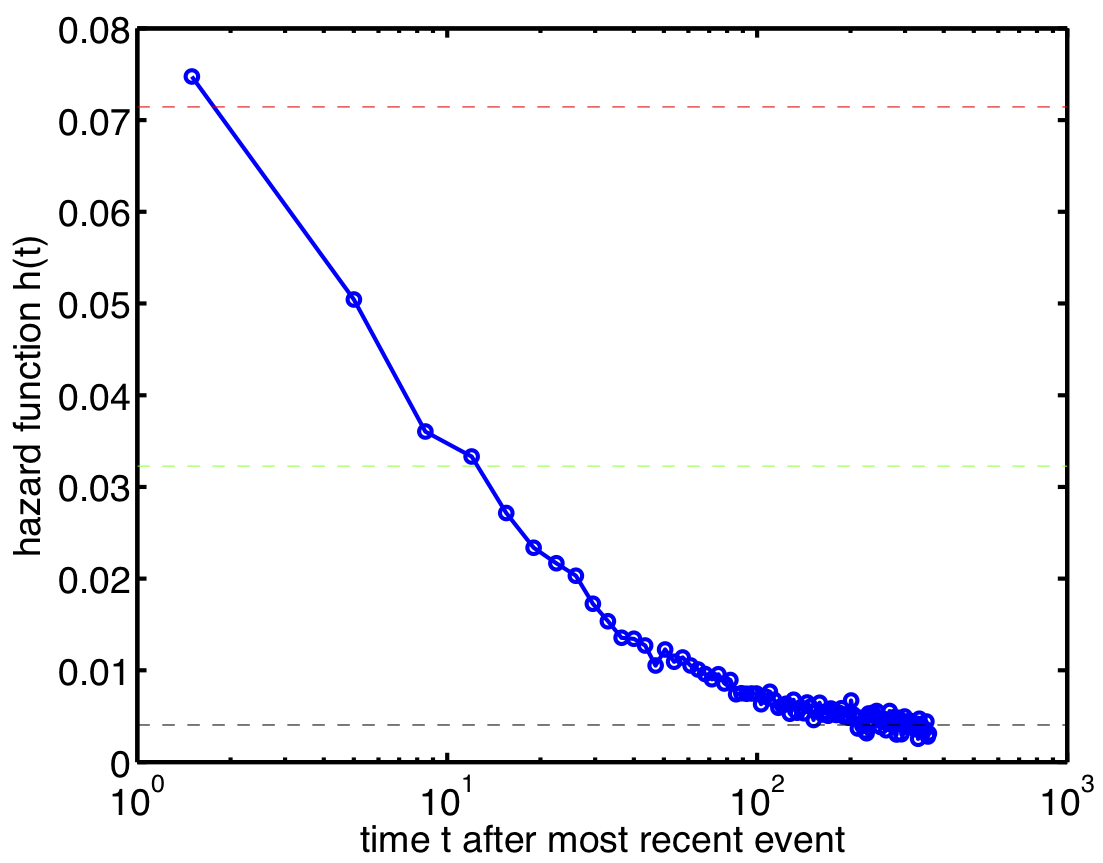} 
    \\[\baselineskip]
    \includegraphics[width=0.49\columnwidth]{%
        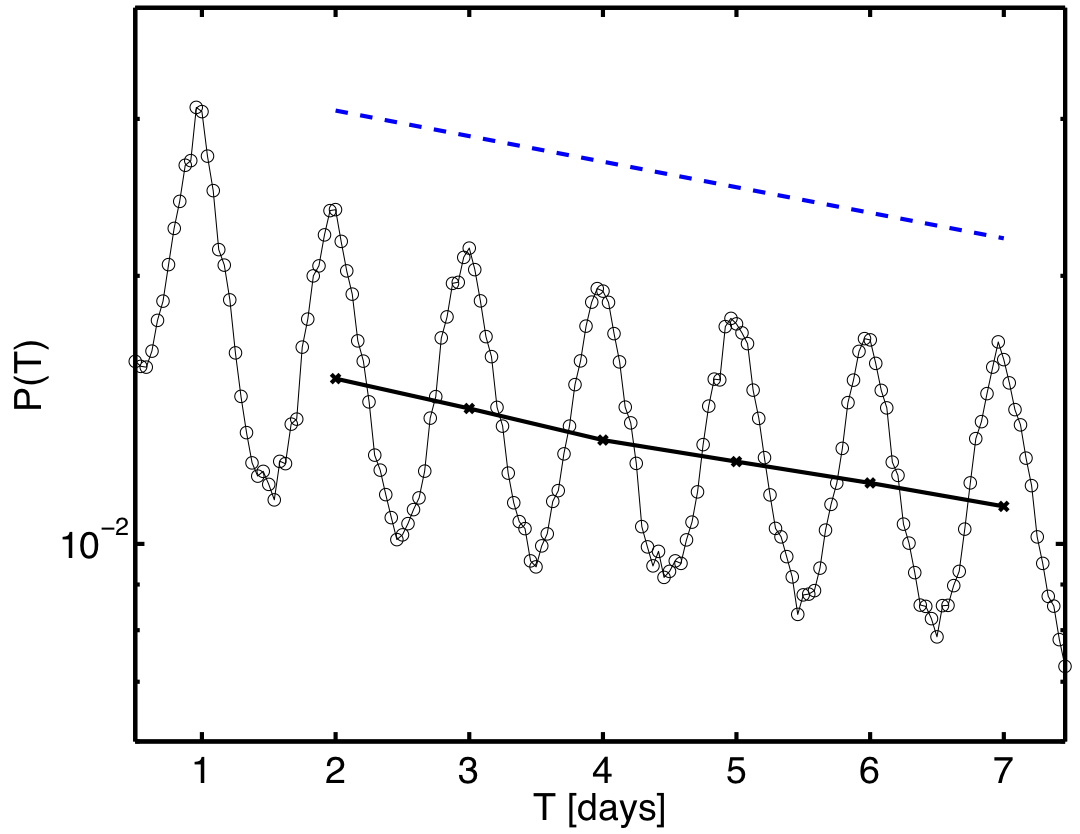}
    \caption{\label{fig:Inter-report-time-statistics.}Inter-report time statistics. (\textbf{Left}) The function $p(t|r<r_{0})$ for $r_{0}=10$\,km. The observed function can be accounted for by an initial algebraic decay $t^{-1}$ moderated by an exponential function for large arguments. The red dashed curved is a fit obtained from maximum-likelihood estimation. (\textbf{Right}) The hazard function $h(t)$ that represents the instantaneous rate of an event at time $t$ provided that an event occurred at $t=0$. The dashed lines represent reporting rates of once per 2 weeks (\emph{top}), once per month (\emph{middle}) and once per $T_{0}=248$~days (\emph{bottom}). (\textbf{Bottom}) $p(t)$ for very short times. A zoom-in resolves daily oscillations modulated by the decay observed on the left. These oscillations indicate that users tend to report to the website at the same time of the day with the highest probability.}
\end{figure}

Complementary to this, temporal aspects of the process can be revealed by computing the pdf for the time $t$ between reporting events given that these occur within a small radius $r>r_{0}$. Figure~\ref{fig:Inter-report-time-statistics.} depicts $p(t)$ for all legs with $r < 10$\,km and a minimal inter-report time of $t_{\text{min}}=1$~day. The inter-report times are described well by a power law moderated by an exponential factor
\begin{equation}
	p(t)\sim t^{-\alpha}e^{-t/T_{0}}\quad\text{with}\quad T_{0}=248\pm27,\quad\alpha=0.99\pm0.05.
	\label{eq:poftfit}
\end{equation}
The observed power-law decay $\sim t^{-1}$ for times $t\ll T_{0}$ is intriguing. These type of decays have been observed in a multitude of contexts involving human activity, for instance the time between consecutive phone calls~\cite{gonzalez2008nature}, emails~\cite{barabasi2005nature} and the number of words between two identical words in texts~\cite{altmann2009plos1}. A consequence of this law is bursting behavior, i.e.~given an event occurred at time $t_{0}$ the probability rate that an event occurs immediately after the first is higher than expected from ordinary Poisson statistics. This behavior is best illustrated by the so-called hazard function $h(t)$ that quantifies the instantaneous probability rate of an event happening at $t$ given that the last event occurred at $t=0$. If we let $$P(\tau>t)=\int_{t}^{\infty}p(s)\,ds$$ be the cumulative probability that the second event occurs at a time $\tau$ later than $t$, the hazard function is defined by $$P(\tau>t)=e^{-\int_{0}^{t}h(s)\,ds}.$$ For a Poisson process with rate $\gamma$ we have $$h(t)=\gamma\quad\Rightarrow\quad P(\tau>t)=e^{-\gamma t}.$$ The hazard function can be computed according to $$h(t)=-\frac{\text{d}}{\text{d}t}\log\left[P(\tau>t)\right]=\frac{p(t)}{P(\tau>t)}.$$ Figure~\ref{fig:Inter-report-time-statistics.} depicts the function $h(t)$ for inter-report times in the WG data. For small times ($t<1$~week) the probability rate for a report is of the order of one report per two weeks, which is also the expected time between two reports in this time window. For larger times ($t>100$~days) the constant value of $1/T_{0}$ is approached, equivalent to one report in 3/4 of a year. Possible explanations of the bursting behavior and the initial algebraic decay in $p(t)$ are a strong behavioral heterogeneity of players that participate in the game or an effective queueing in the system, i.e.~bills may enter shops and initially have a comparatively high likelihood of leaving, being {}``on top of the stack''. As time passes these bills may {}``get stuck'' and equilibrate to the long time scale present in the system.

\subsection{Definition of the mobility network}

\begin{figure}
    \sidecaption
    \includegraphics[width=6cm]{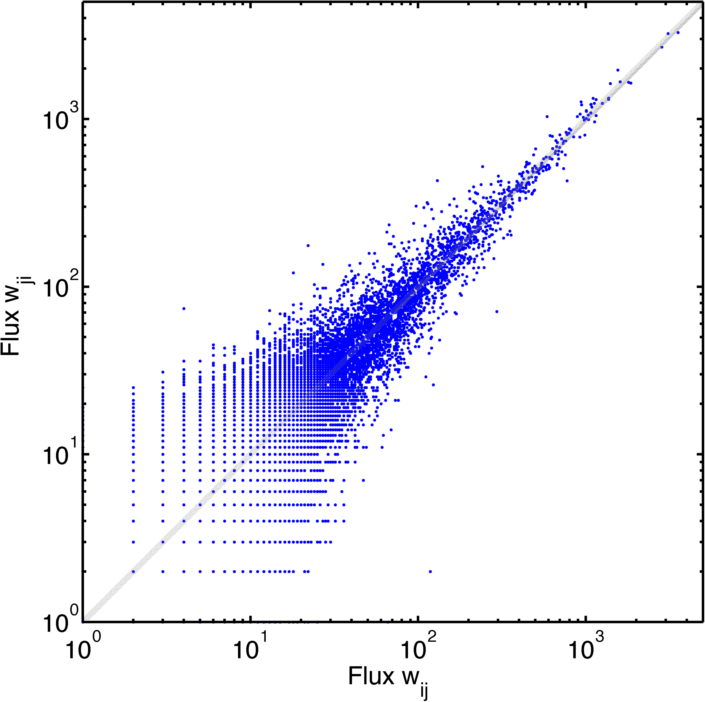}
    \caption{\label{fig:Symmetry-of-flux}Symmetry of flux network $\tilde{w}_{ij}$.}
\end{figure}

\noindent From the trajectories defined by~(\ref{eq:noIEtrajectories}) and the characteristic functions of the counties~\eqref{eq:countyPoly} we construct a matrix $\tilde{w}_{ij}$ that counts the number of legs which originate at county $i$ and terminate at $j$,
$$\tilde{w}_{ij}=\sum_{n=1}^N\,\sum_{k=2}^{L_{n}}\,\chi_{i}\left(\mathbf{X}_{n,k-1}\right)\,\chi_{j}\left(\mathbf{X}_{n,k}\right)\,\Theta(T-\Delta T_{n,k})$$
where $\Theta(\cdot)$ is the Heaviside step-function. In order to exclude potential biases induced by initial entries we ignore the first leg of all trajectories ($k=2$ in the above sum). This choice is motivated by the fact that the community of individuals that initiate bills might be less representative than those that find bills and report them. Indications that this might have an effect are supported by the different scaling behavior of initial entry frequencies with population as compared to report frequencies with population. The factor $\Theta(T-\Delta T_{n,k})$ excludes legs that have an inter-event time larger than time $T$. The matrix $\tilde{w}_{ij}$ need not to be symmetric, as the flux of bills from $i\rightarrow j$ need not equal those that travel $j\rightarrow i$. However, as Figure~\ref{fig:Symmetry-of-flux} indicates the flux matrix is statistically symmetric. Plotting $\tilde{w}_{ij}$ against $\tilde{w}_{ji}$ indicates a clear mean linear relationship. Since we base our analysis on the flux of money between two given counties we symmetrize the network and use $w_{ij}$ in our analysis defined by $$w_{ij}=\frac{1}{2}\left(\tilde{w}_{ij}+\tilde{w}_{ji}\right)$$ which of course also depends on the time threshold parameter $T$. Choosing the optimal value for $T$ is a trade-off between trying to estimate instantaneous flux, i.e.~choosing $T$ as small as possible, and using as many legs as possible to decrease fluctuations, i.e.~choosing large values for $T$. Choosing a value for $T<30$~days for instance rules out bills that visit a Federal Reserve Bank in between reports in counties $i$ and $j$, as bills that enter FRBs do not return to circulation until approximately 3--4 weeks after entering the FRB. To make sure that our results do not significantly change as the parameter $T$ is varied we performed the analysis for various values of $T$ ranging from a few days to $T=1$~year. The computed border structure does not significantly depend on the value of $T$. Decreasing $T$ thins out the network and reduces the overall connectivity, yet the effects are similar to bootstrapping the network randomly, a process that also does not change our results and is discussed in Section~\ref{sec:boot}.

\subsection{Gravity as a null model} \label{sec:grav_construction}

% TODO We should be more specific about the Census reference
In addition to the empirical data described above, we also construct a synthetic mobility network based on the well-known gravity law hypothesis~\cite{Anderson:1979p1765, Cochrane:1975p1767, Xia:2004p1775} to serve as a null model. In gravity models the interaction strength between a collection of sub-populations with geographic positions $x_i$, sizes $N_i$ (obtained from census data\footnote{\url{http://www.census.gov}}), and distances $d_{ij} = \left|x_i - x_j\right|$ is given by
\begin{equation}\label{eq:grav}
p_{ij}\propto \frac{N_i^\alpha N_j^\beta}{{d_{ij}}^{1+\mu}}
\end{equation}
in which $\alpha$, $\beta$ and $\mu$ are non-negative parameters.

To create a model network comparable to our data, we first compute $p_{ij}$ for all counties $i$ and $j$ in the continental U.S. and normalize them such that $\sum_{i,j} p_{ij} = 1$.  We then interpret these values as probabilities for a travel event to happen between the two counties (or, speaking in terms of the original data source, a dollar bill report).  Thus, starting with all-zero link weights $w_{ij}$, we repeatedly draw a pair of nodes according to $p_{ij}$ and increase the corresponding $w_{ij}$ by one, until approximately the same connectivity (number of non-zero $w_{ij}$) as in the real-data network is reached.

We generated gravity networks for different parameter values and gauged them against our real data by comparing the distributions of first-order network statistics to find the best fit to our data.  Distributions have been compared by log-binning the values and computing the $\chi^2$ statistic $$\chi^2 = \sum_i^n \frac{(N_i^G - N_i^R)^2}{N_i^R}$$ where $n$ is the number of bins and $N_i^G$ ($N_i^R$) is the number of values from the gravity (real-data) network in bin $i$.  

\begin{figure}
    \sidecaption
    \includegraphics[width=7cm]{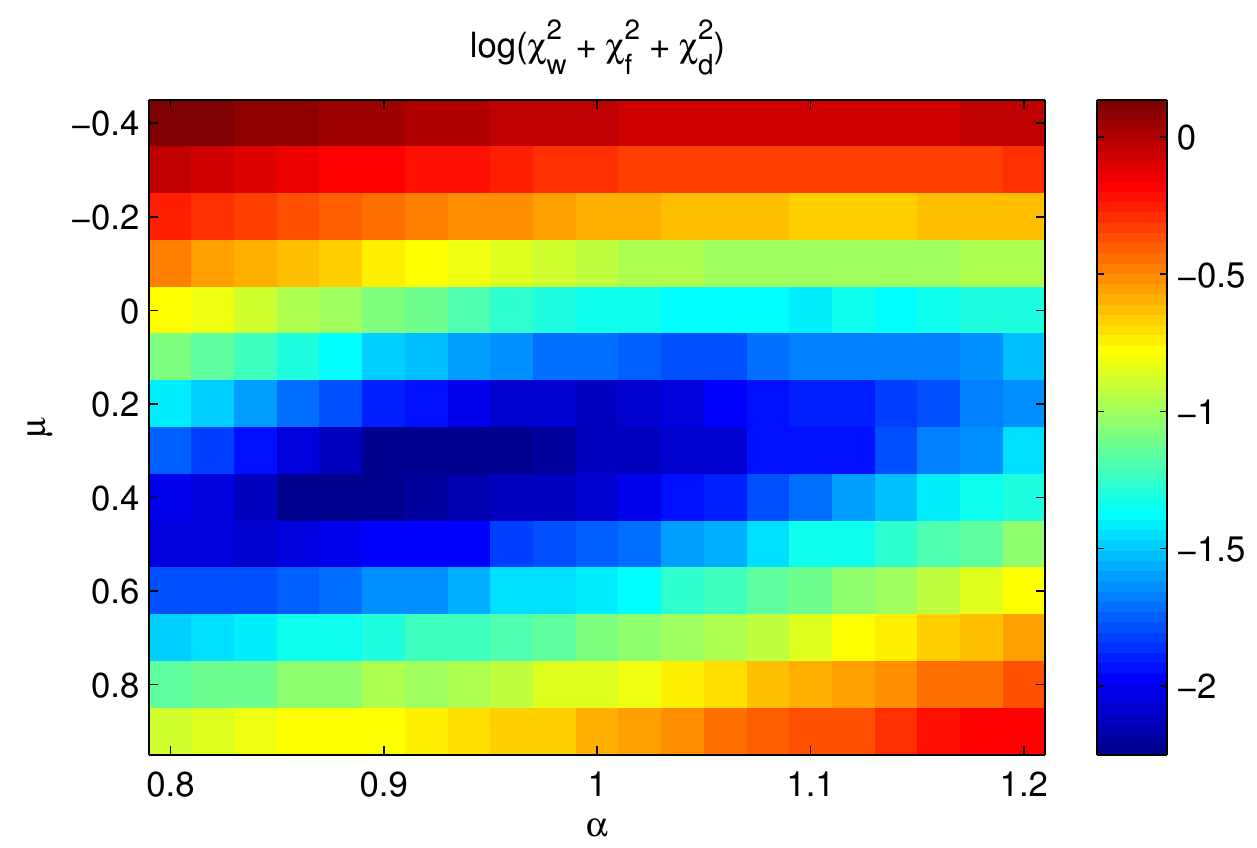}
    \caption{\label{fig:chi2fit}$\chi^2$ goodness-of-fit for different parameters of the gravity law.  The minimum is at $(\alpha,\mu) = (0.96,0.3)$.}
\end{figure}

\begin{figure}
    \centering
    \includegraphics[width=.75\linewidth]{% 
        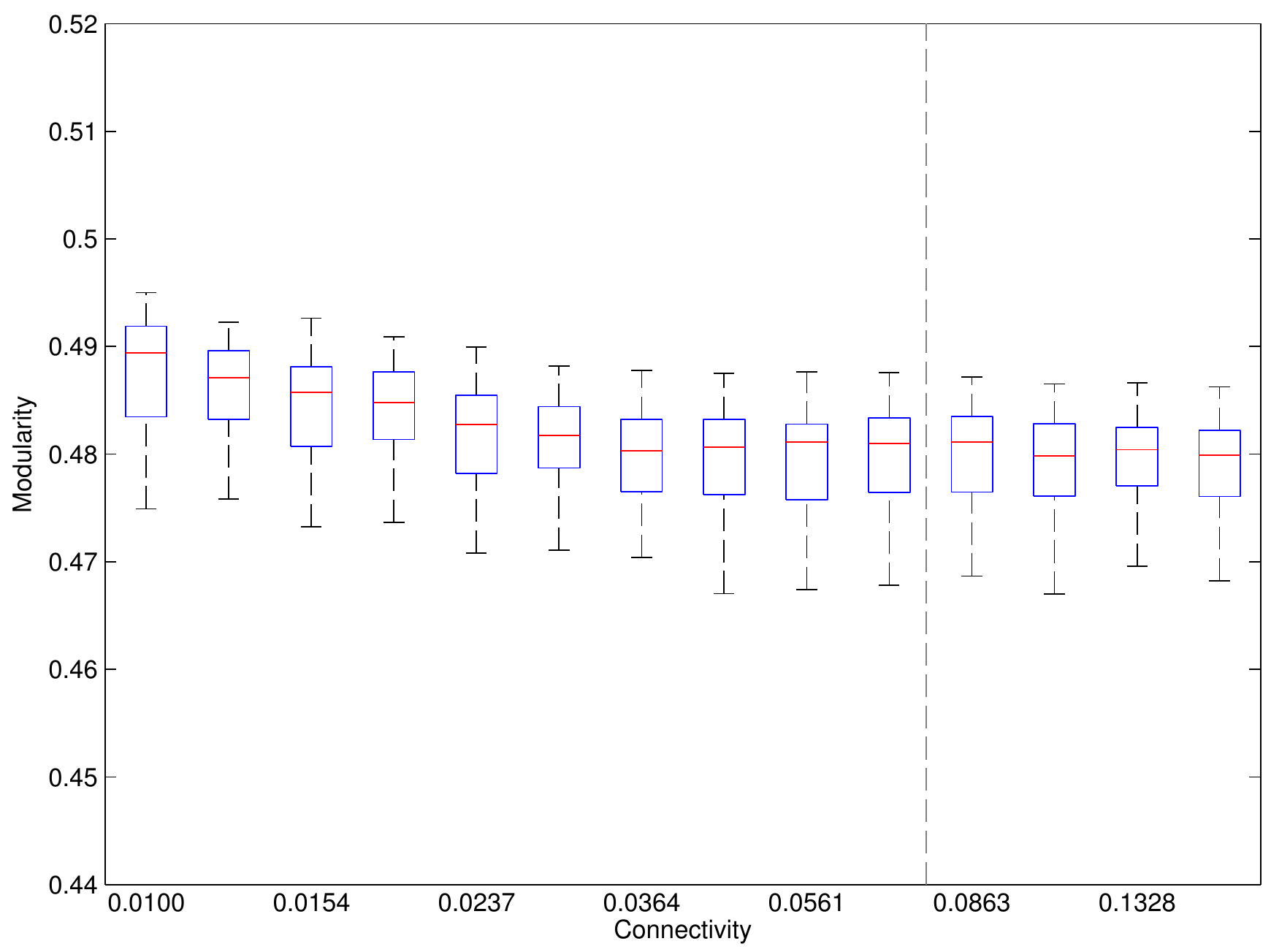}
    \caption{\label{fig:bootstrap_mods} Distributions of modularity values for an ensemble of 80~partitions each computed for snapshots of the model network at different connectivities.  The dashed line corresponds to 0.0765, the connectivity of the real-data mobility network.}
\end{figure}

\begin{figure}
    \centering\mbox{}\hfill
    \includegraphics[width=.4\linewidth]{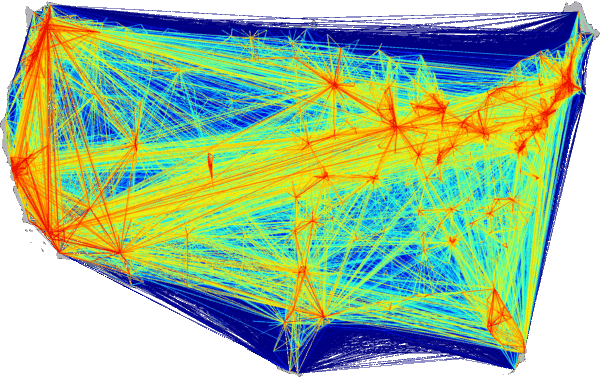}\hfill
    \includegraphics[width=.4\linewidth]{% 
        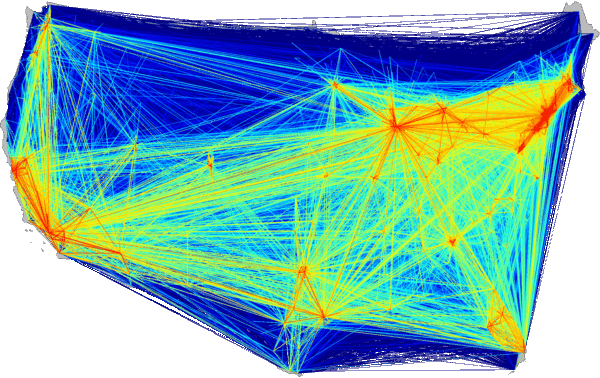}\hfill\mbox{}\\[\baselineskip]
    \includegraphics[width=.85\linewidth]{%
        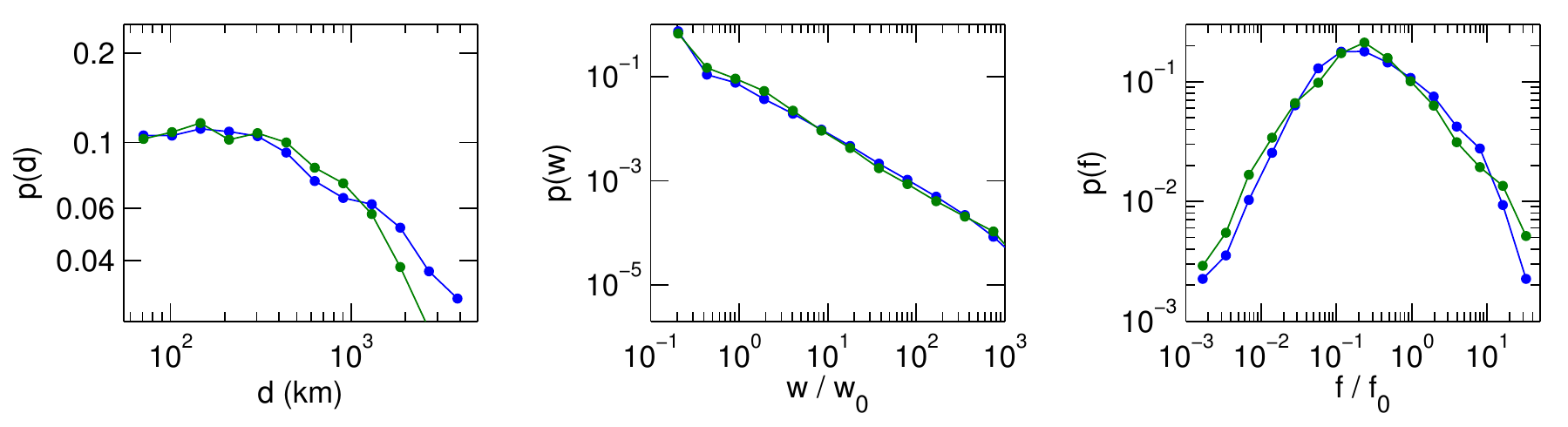}
    \caption{\label{fig:grav}Comparison of the real-data network (top left) and the gravity model network with $\alpha = \beta = 0.96$ and $\mu = 0.3$ (top right). The bottom plot shows the distributions of geographical distances $d$, link weights $w$, and node fluxes $f$ in the real-data network (blue lines) and the gravity model (green lines).}
\end{figure}

Our real data is symmetric and node fluxes are proportional to population sizes, therefore we assume $\alpha = \beta\approx 1$ to narrow down the search volume in parameter space.  We computed $\chi^2$ for the distribution of link weights, node fluxes and geographical distances and used the sum of them, $\chi^2_w + \chi^2_f + \chi^2_d$, as the goodness-of-it measure.  Figure~\ref{fig:chi2fit} shows this quantity for $(\alpha,\mu)\in[0.8,1.2]\times[-0.4,0.9]$, from which we concluded that $\alpha = \beta = 0.96$ and $\mu = 0.3$ are the best parameter choices.  The resulting network and first-order statistics are shown in Figure~\ref{fig:grav}.

Similar to the bootstrapping procedure described in Section~\ref{sec:boot}, we tested the robustness of the community structure of the model network by generating snapshots of the network at different connectivities and computing an ensemble of 80~high-modularity partitions for each snapshot.  We found that the modularity statistics are stable around the target connectivity of 0.0765 (Figure~\ref{fig:bootstrap_mods}).

\section{Degeneracy and superposition} \label{sec:degeneracy}

\begin{figure}
    \mbox{}\hfill
    \includegraphics[width=.45\linewidth]% 
        {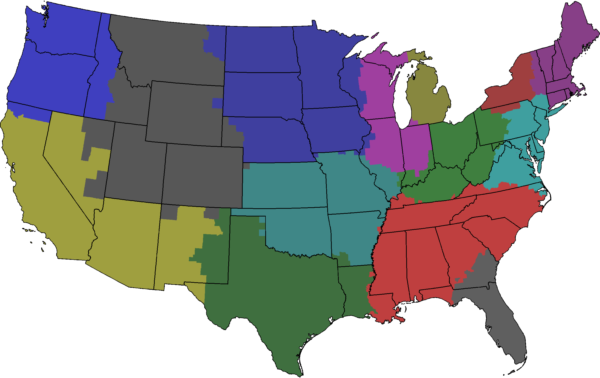} \hfill
    \includegraphics[width=.45\linewidth]% 
        {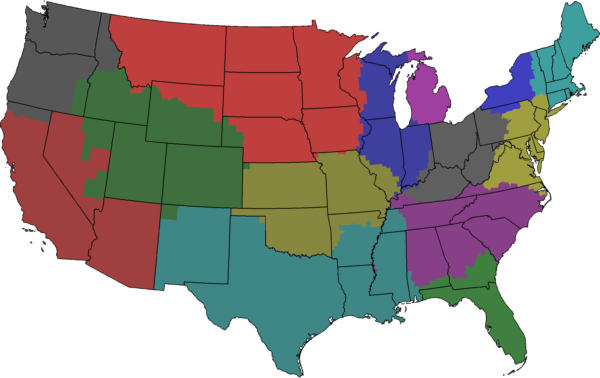}
    \mbox{}\hfill\\
    \caption{High-modularity community partitions of the WG mobility network. The stochastic algorithm produces different partitions when run many times; these are two representative examples. Modularity values are 0.6808 (left) and 0.6807 (right).}
    \label{fig:samplepartitions}
\end{figure}

Given a mobility network constructed from the Where's George data, we then apply the optimization algorithm described in Section~\ref{sec:modmax} to generate community partitions. Since the optimization process is stochastic, the resulting partition varies between realizations of the process. Two representative examples of high-modularity partitions are displayed in Figure~\ref{fig:samplepartitions}. Note that, although modularity only takes into account the structure of the weight matrix $W$ and is explicitly blind to the geographic locations of nodes, the effective large-scale modules are spatially compact in every map. Consequently, although long-distance mobility plays an important role, the massive traffic along short distances generates spatial coherence of community patches of mean linear extension $l=633 \pm 250$\,km. Note however that although each maps exhibits qualitative similarities between detected large-scale subdivisions and although each of the maps possess a high modularity score, obvious structural differences exist; in fact, even if it were possible to determine a partition with maximal modularity, any such partition is not in principle unique. It is thus questionable whether any single effective map can be considered the most plausible partition.

Theoretical concerns aside, recent work~\cite{Fortunato:2007p1193} has identified practical issues with modularity maximization, in particular the so-called \emph{resolution limit}. We demonstrate that a superposition of community partitions can alleviate these issues with the modularity score, and to this end discuss its known shortcomings in more detail.

In fact, it is straightforward to construct networks of which several distinct partitions with equal and maximum modularity value exist. This \emph{degeneracy of modularity} was independently found by Good~et~al.~\cite{good2009arxiv} and marked as a drawback of the modularity measure.

\begin{figure}
    \centering\includegraphics[width=.85\linewidth]{% 
        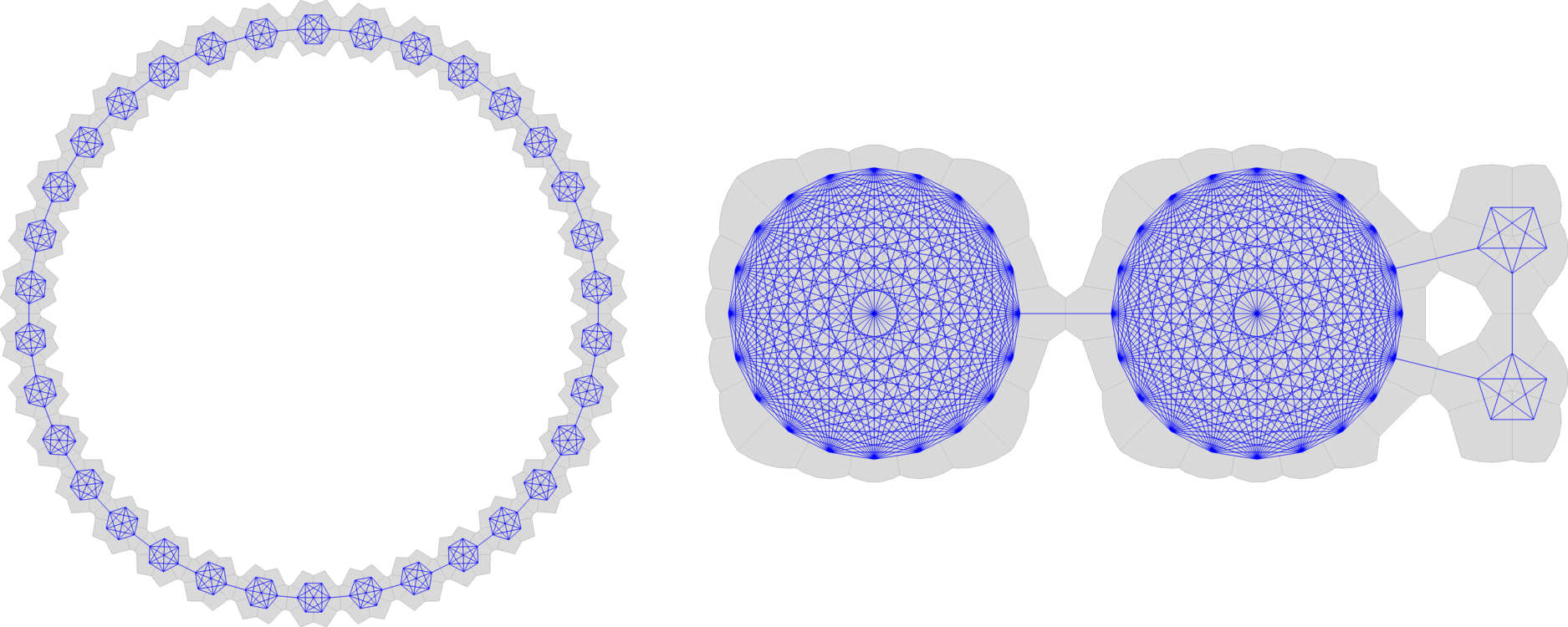}
    \caption{\label{fig:reslim_networks}Two networks that expose the resolution limit problem with modularity. The shaded areas indicate an artificial geography for nicer visualization of the boundaries in the next figures.  (\textbf{Left}) A ring of 34 cliques, each of 6 nodes and connected to their neighbors by single links.  (\textbf{Right}) A network of two 20-node cliques and two five-node cliques.}
\end{figure}

Fortunato and Barth{\'e}lemy~\cite{Fortunato:2007p1193} also report on the resolution limit of modularity. The authors present two artificial, unweighted networks that exhibit an intuitively very clear community structure, yet partitions exist that do not reflect this structure but have a higher modularity value than the partition that does. In particular, these networks are constructed by connecting multiple fully connected graphs (``cliques'') with single links (Figure~\ref{fig:reslim_networks}). It is clear that every clique should be grouped into one module, but the best partition according to modularity will group multiple cliques together. This only occurs if the cliques are small (in terms of number of links) compared to the full network, thus the modularity measure cannot detect communities below a certain resolution limit.

Our proposed method combines an ensemble of partitions by focusing on the boundaries of a partition (``Which adjacent nodes are separated into different modules?'') rather than its volumes (``Which nodes are grouped together?''), and then computing for each boundary the fraction of partitions in which it exists. Because we are interested in geographically embedded networks and modules are virtually always spatially compact in our case, we can restrict ourselves to boundaries that are also real geographical borders between nodes. However, the idea can be easily generalized to non-geographical networks, at the expense of convenient straight-forward visualization. Since all partitions in the ensemble have a high modularity value, this method highlights similarities and differences in degenerated partitions, yielding a unique ``partition'' (or to be more precise, a map) of the network and thus overcoming the degeneracy problem.

\begin{figure}
    \centering\includegraphics[width=\linewidth]{%
        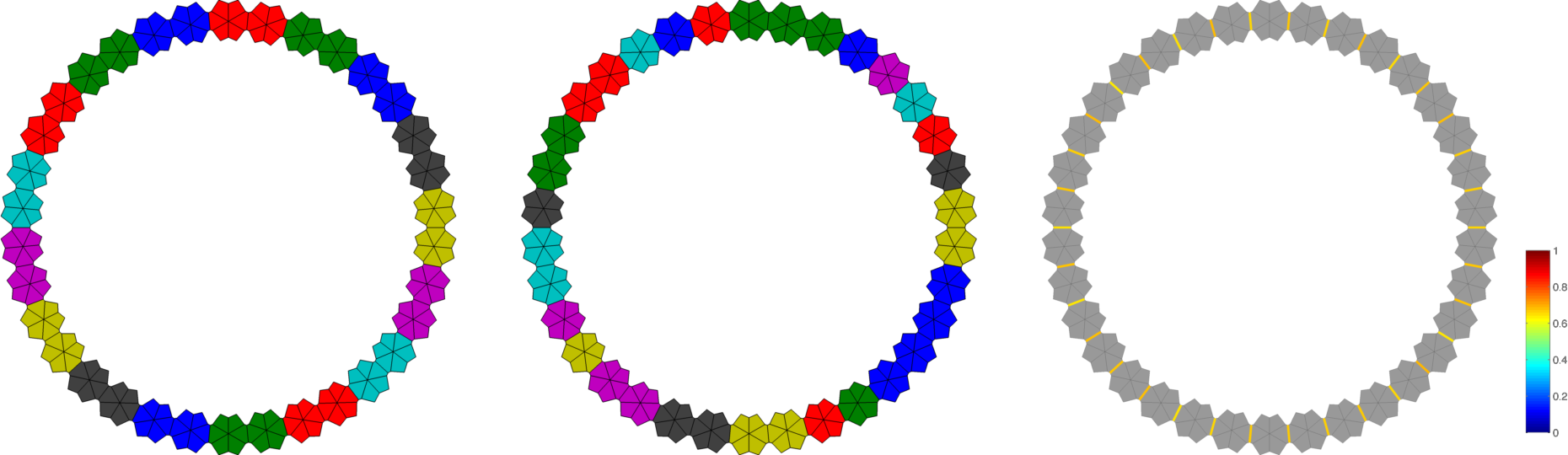}
    \caption{\label{fig:wicked_clique_ring}(\textbf{Left}) The optimal partition in the clique ring groups pairs of cliques together (the same color is used for multiple modules).  (\textbf{Center}) Example of a partition found by the modularity optimization algorithm.  (\textbf{Right}) Superposition reveals boundaries in the clique ring between every clique. Color codes the fraction of partitions in which the boundary was found. We use $T=2.5 \cdot 10^{-4}$, $c=0.75$, and $f=0.5$ for this example and the next.}
\end{figure}

In our method any single partition obviously suffers from this limitation as well. However, the resolution limit can be alleviated by looking at an ensemble, if enough small modules exist to create degeneracies. To illustrate this, we applied our method to the two example networks from Fortunato and Barth\'elemy~\cite{Fortunato:2007p1193}. Figure~\ref{fig:reslim_networks} (left) shows a ring of 34 6-cliques, all connected to their neighbors by a single link. The intuitive partition in which each clique is in its own module has modularity $Q_\text{real} = 0.9081$ while a partition that groups pairs of cliques together has $Q_\text{opt} = 0.9099$. However, two distinct partitions exist that group pairs of cliques. Thus, an ensemble of optimal partitions will be composed out of those two partitions, yielding a boundary map in which every boundary between two cliques appears in 50\% of the ensemble partitions. For nicer visualization, we created an artificial geography for this network and computed partitions and boundaries, shown in Figure~\ref{fig:wicked_clique_ring}. Due to the nature of our algorithm, the resulting partitions contain a few $n$-tuples of cliques and single-clique modules that have not been split or merged into perfect clique-pairs before the termination criterion, and thus the observed boundaries are stronger than expected.

\begin{figure}
    \centering\includegraphics[width=0.9\linewidth]{%
        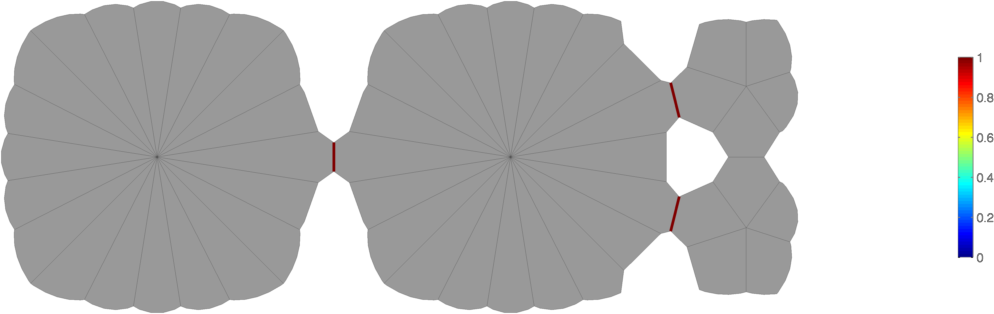}
    \caption{\label{fig:wicked_clique_group}Boundaries found in the clique network shown in Figure~\ref{fig:reslim_networks} (right).  Our algorithm is not able to find a boundary between the two small cliques.}
\end{figure}

The second network proposed in Fortunato and Barth\'elemy is constructed from two 20-cliques and two 5-cliques (Figure~\ref{fig:reslim_networks} (right)).  Here, the two smaller cliques are merged into one module by the optimal partition ($Q_\text{opt} = 0.5426$), although one would again expect each of them to be in its own module ($Q_\text{real} = 0.5416$).  Our method is not able to capture the intuitive community structure in this case (Figure~\ref{fig:wicked_clique_group}), because no degeneracy exists (the partitions in which only one of the small cliques are grouped with the large one, but not the other, are too far from the optimum to be produced by the algorithm, $Q_{deg} = 0.4959$).

\begin{figure}
    \centering
    \includegraphics[width=0.83\linewidth]{%
        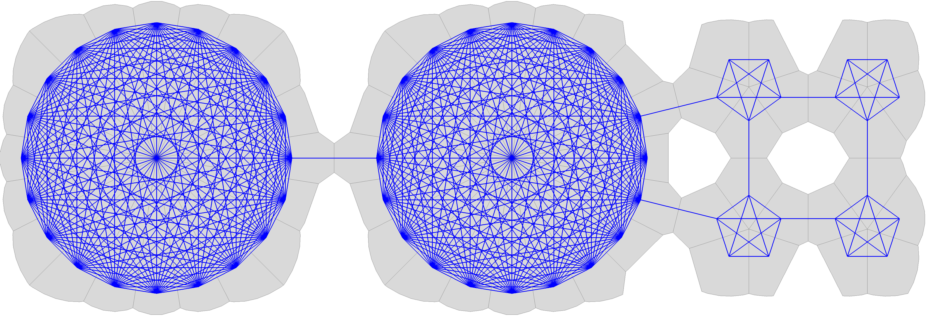}\\[\baselineskip]
    \hspace{17pt}
    \includegraphics[width=0.9\linewidth]{%
        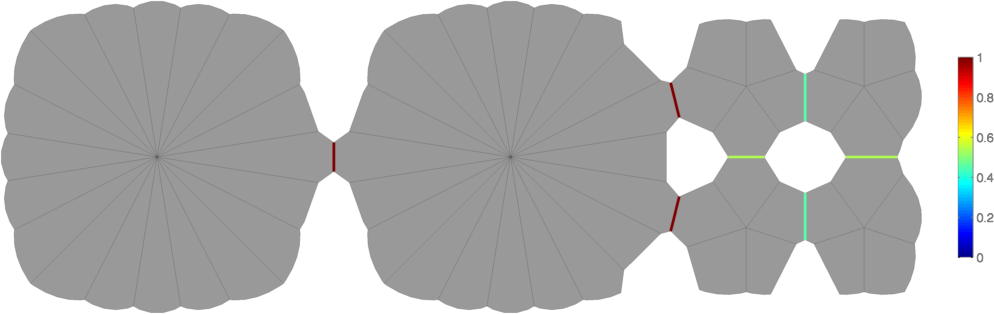}
    \caption{\label{fig:wicked_clique_group_mod3}Modification of the clique network in Figure~\ref{fig:reslim_networks} (right).  Because there are multiple high-modularity partitions that group the smaller cliques into pairs, our method can detect the correct community structure in this case.}
\end{figure}

But if we extend the network such that four small cliques exist, the partition which groups all cliques into their own modules is still suboptimal to any partition that groups together more than one of the small cliques, but degeneracies are created and the ensemble of partitions reveals the true community structure in this network (Figure~\ref{fig:wicked_clique_group_mod3}).

In conclusion, our method is able to dissolve both the degeneracy and resolution limit problems if enough small modules exist to create degeneracies.  In fact, we will observe small ``building blocks'' in the WG data that are not seen in single partitions but emerge from the superposition of a partition ensemble.

\section{Assessment of border structures} \label{sec:borders}

\begin{figure}
    \sidecaption
    \includegraphics[width=6cm]{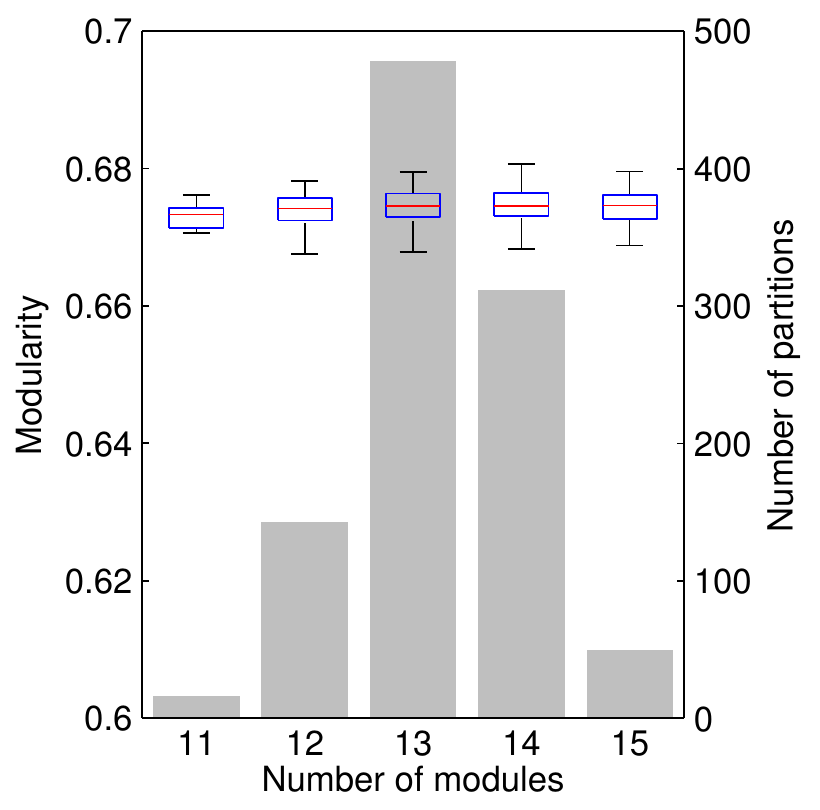}
    \caption{Ensemble statistics of geographic subdivisions for a set of $N=1,000$ partitions. The number of modules $k$ in each subdivision is narrowly distributed around 13 (grey bars), and so are the conditional distributions of modularity (superimposed whisker plots). The ensemble mean is $\overline{Q} = 0.674 \pm 0.0026$.}
    \label{fig:ensemble_mod_stats}
\end{figure}

\begin{figure}
    \centering
    \includegraphics[width=\textwidth]%
        {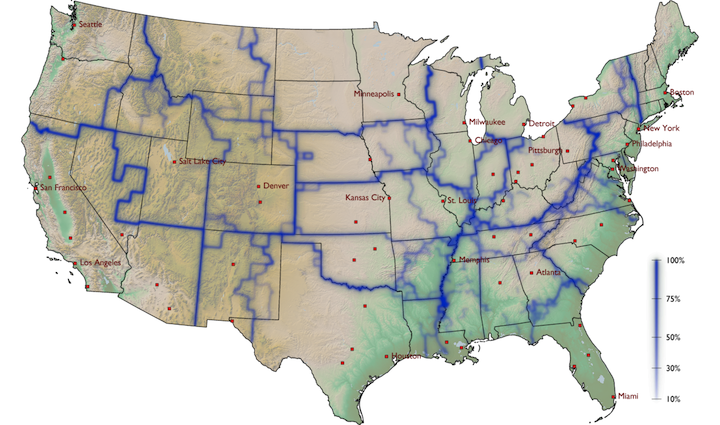}
    \caption{Effective borders emerge from linear superposition of all maps in the ensemble (blue lines). Intensity encodes border significance (i.e.\ the fraction of maps that exhibit the border). Black lines indicate state borders. Although 44\% of state borders coincide with effective borders (left pie chart), approximately 64\% of effective borders do not coincide with state borders. These borders are statistically significant features of the ensemble of high modularity maps, they partially correlate with administrative borders, topographical features, and frequently split states.}
    \label{fig:border_stucture}
\end{figure}

Using the algorithm described in Section~\ref{sec:modmax}, we compute an ensemble of 1,000~partitions of the WG mobility network, all exhibiting a high modularity ($Q = 0.6744 \pm 0.0026$, see also Figure~\ref{fig:ensemble_mod_stats} for the distribution of modularity values) and spatially compact modules, and perform a linear superposition of the set of maps. This method extracts features that are structural properties of the entire ensemble. The most prominent emergent feature is a complex network of spatially continuous geographic borders (Figure~\ref{fig:border_stucture}). These borders are statistically significant topological features of the underlying multi-scale mobility network. An important aspect of this method is the ability to not only identify the location of these borders but also to quantify the frequency with which individual borders appear in the set of partitions, a measure for the strength of a border.

Investigating this system of effective mobility borders more closely, we see that although they correlate significantly with territorial state borders ($p<0.001$, see Section~\ref{sec:comparisons}) they frequently occur in unexpected locations. For example, they effectively split some states into independent patches, as with Pennsylvania, where the strongest border of the map separates the state into regions centered around Pittsburgh and Philadelphia. Other examples are Missouri, which is split into two halves, the eastern part dominated by St.~Louis (also taking a piece of Illinois) and the western by Kansas City, and the southern part of Georgia, which is effectively allocated to Florida. Also of note are the Appalachian mountains. Representing a real topographical barrier to most means of transportation, this mountain range only partially coincides with state borders, but the effective mobility border is clearly correlated with it. Finally, note that effective patches are often centered around large metropolitan areas that represent hubs in the transportation network, for instance Atlanta, Minneapolis and Salt Lake City. We find that 44\% of the administrative state borders are also effective boundaries, while 64\% of all effective boundaries do not coincide with state borders.

\subsection{Comparison to gravity models}
\begin{figure}
    \centering
    \includegraphics[width=\textwidth]% 
        {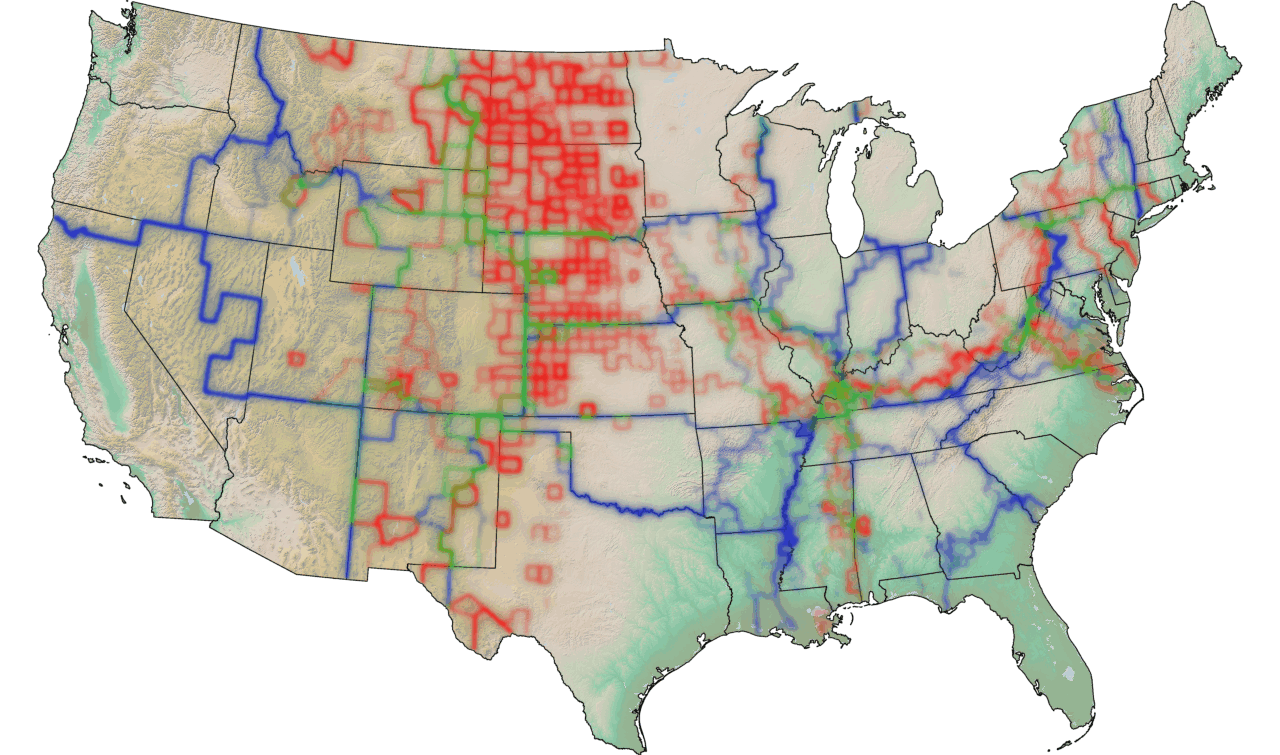}
    \caption{The border structure of the gravity network (red) partially coincides with the borders in the original data (blue), but not significantly.  The overlap is shown in green, for significance tests see Section~\ref{sec:comparisons}.}
    \label{fig:grav_border_structure}
\end{figure}

We also investigate whether the observed pattern of borders can be accounted for by the prominent class of gravity models~\cite{Anderson:1979p1765, Cochrane:1975p1767, Xia:2004p1775}, frequently encountered in modeling spatial disease dynamics~\cite{Xia:2004p1775}. In these phenomenological models it is assumed that the interaction strength $w_{ij}$ between a collection of sub-populations is given by \eqref{eq:grav}, and we construct such a model according to the procedure described in Section~\ref{sec:grav_construction}. Although their validity is still a matter of debate, gravity models are commonly used if no direct data on mobility is available. The key feature of a gravity model is that $w_{ij}$ is entirely determined by the spatial distribution of sub-populations. We therefore test whether the observed patterns of borders (Figure~\ref{fig:border_stucture}) are indeed determined by the existing multi-scale mobility network or rather indirectly by the underlying spatial distribution of the population in combination with gravity law coupling. Figure~\ref{fig:grav_border_structure} illustrates the borders we find in a network that obeys equation~(\ref{eq:grav}).

\begin{figure}
    \centering
    \includegraphics[width=0.45\textwidth]% 
        {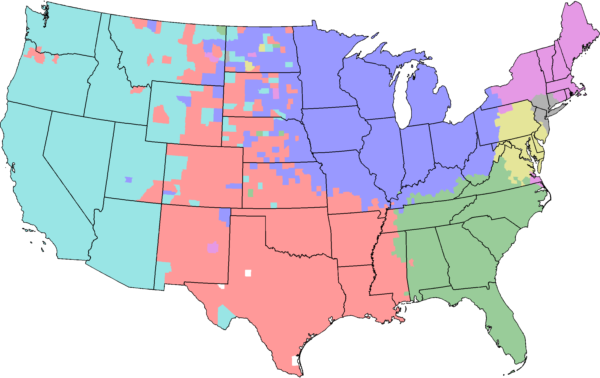} \hfill
    \includegraphics[width=0.45\textwidth]% 
        {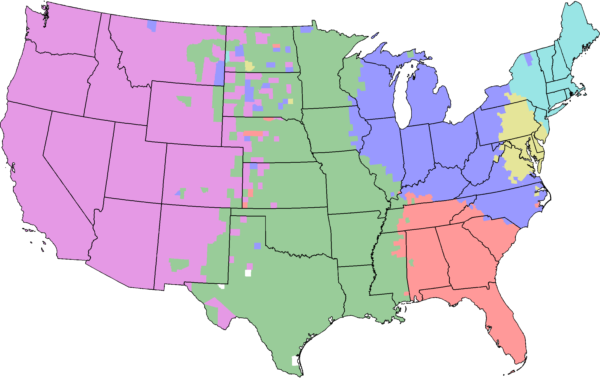}
    \caption{Sample partitions of the gravity network. Although they share qualitative features with those from the original network (Figure~\ref{fig:samplepartitions}), generic partitions of the gravity model network are structurally different, typically exhibiting fewer modules per partition, in different locations and with less spatial compactness.}
    \label{fig:grav_sample_partitions}
\end{figure}

Comparing this model network to the original multi-scale network we see that their qualitative properties are similar, with strong short-range connections as well as prominent long-range links. However, maximal modularity maps typically contain only five subdivisions with a mean modularity of only $\bar Q= 0.4791$. Because borders determined for the model system are strongly fluctuating (Figure~\ref{fig:grav_sample_partitions}), they yield much less coherent large-scale patches. Some specific borders, e.g.\ the Appalachian rim, are correctly reproduced in the model. The difference between the borders of the model system and the empirical data is statistically significant (see Section~\ref{sec:comparisons}), and we conclude that the sharp definition of borders in the original multi-scale mobility network and the pronounced spatial coherence of the building blocks are an intrinsic feature of the real multi-scale mobility network and cannot be generated by a gravity model that has a maximum first-order statistical overlap with the original mobility network.

\section{Shortest-path tree clustering} % (fold)
\begin{figure}
    \centering
    \includegraphics[type=png,ext=.png,read=.png,width=0.9\textwidth]% 
        {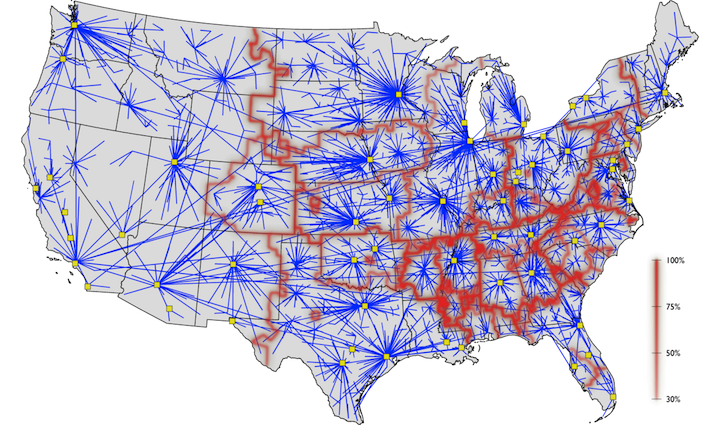}
    \caption{By comparing the border structure from SPT clustering with the ensemble of significant links (those that appear in at least half of the shortest-path trees) we identify topological structures which reveal the core of the network that explains the majority of border locations. This core is represented by the network in blue consisting of star-shaped modules centered around large cities (yellow squares).}
    \label{fig:boundarymapspt}
\end{figure}

\label{sec:shortest_path_tree_clustering}
The methods already discussed successfully extract the structure of geographic borders inherent in multi-scale mobility networks. Bootstrapping the network indicates that these structures are surprisingly stable in response to perturbations of the network, but neither the modularity measure nor the stochastic algorithm we use to discover partitions provide specific information about the substructures in the network that make these borders so robust. What feature of the network, more specifically which subset of links if any, generates the observed borders? In order to address this question and further investigate the structural stability of the observed patterns, we developed a new and efficient computational technique based on the concept of shortest-path trees~(SPT). Like stochastic modularity maximization, this technique identifies a structure of borders that encompass spatially coherent regions (Figure~\ref{fig:boundarymapspt}), but unlike modularity this structure is unique. More importantly, it identifies a unique set of connections in the network, a network backbone, that correlates strongly with the observed borders.

This second method for identifying community partitions, based on topological features of the analyzed network, has three parts. Given a network with $N$ nodes containing a single connected component, we first compute a shortest-path tree for each node in the network. At least three widely-known algorithms are applicable (Dijkstra, Floyd-Warshall, and Bellman-Ford) and various optimizations are possible; in addition, if the input is sparse some of these algorithms improve in time complexity. In the worst case, however, this can be computed in $O(N^3)$ time.

Second, we compute a dissimilarity score for each pair of shortest-path trees, and using the dissimilarity functions described below, this can also be accomplished in $O(N^3)$ time.

Third and last, we apply hierarchical clustering to the table of dissimilarity scores, which also takes $O(N^3)$ time for a naive implementation (because we compute the smallest element of an at-largest-$N$-by-$N$ table $N$ times). Therefore the entire suggested procedure takes $O(N^3)$ time.

As mentioned, various optimizations are possible for computing shortest-path trees and hierarchical clustering, and these algorithms are so widely used that high-quality, efficient implementations are easily available. In fact, we find that the second step, computing dissimilarity scores, actually dominates the running time although it is by far the simplest computation; this is due to the fact that we use interfaces to pre-compiled, canned routines for steps 1 and 3, while step 2 is a naive MATLAB script. In practice the entire analysis can be run start to finish in under a half hour for our network of $N=3,109$ nodes on a circa-2008 laptop.

\subsection{Computing shortest-path trees} % (fold)
\label{sub:computing_shortest_path_trees}
The shortest path from vertex $i$ to vertex $j$ is the series of edges that minimizes the total effective distance $d=\sum 1/w_{ij}$ along the legs of the path\cite{Dijkstra:1959p1654}. The distance along an edge for us is the inverse of the edge weight, as a highly-weighted edge indicates that two vertices are effectively proximal. (There are no edges with an infinite distance, because we do not define an edge between vertices if there is zero weight.)

The shortest-path tree $T_i$ rooted at node $i$ is the union of all shortest paths originating at $i$ and ending at other nodes. We use the MATLAB interface\footnote{\url{http://www.stanford.edu/~dgleich/programs/matlab_bgl/}} to the Boost Graph Library\footnote{\url{http://www.boost.org/doc/libs/1_41_0/libs/graph/doc/index.html}} to compute shortest-path trees. To prevent random fluctuations in our data from overwhelming the signal, we add a weak link between neighboring counties.

% subsection computing_shortest_path_trees (end)

\subsection{Measuring tree distance} % (fold)
\label{sub:measuring_tree_distance}
A shortest-path tree can be easily represented as a vector of vertex labels $T = [t_k], k = 1 \dots N$, such that $t_k$ is the label of the parent of vertex $k$, with a special symbol (perhaps $0$) used to indicate the root. There are no disconnected nodes in the mobility network, thus each tree vector represents a single tree and not a forest. This representation lends itself to straightforward and meaningful comparisons between two trees.

We define two related measures of the dissimilarity between two trees. The first, called \emph{parent dissimilarity}, asks the question, how many of the vertices in $T_A$ do not have the same parent in $T_B$? We denote this by $z_p(T_A,T_B)$, and it is exactly the general Hamming distance of two symbol sequences, that is, the number of places where corresponding labels in $T_A$ and $T_B$ do not match. The second, called \emph{overlap dissimilarity}, asks the question, how many edges do the two trees \emph{not} share? It is defined as $z_o(T_A,T_B) = s_{max} - s(T_A,T_B)$. Here, $s_{max}$ is the largest number of edges two trees could share, which is the number of vertices less one (since the root does not contribute an edge). $s(T_A,T_B)$ is the number of edges that $T_A$ and $T_B$ \emph{do} share, and where $z_p$ asks essentially the same question considering edges to be directed, $z_o$ considers edges to be undirected. Also note that although we consider only the topology of trees when measuring their dissimilarity, the topology is determined by the weight of edges in the original graph and thus the mobility dynamics. For both measures, possible $z$ values range from $0$ (completely identical trees) to $N$, the number of nodes in the network.

We compute both measures for each distinct pair of trees in our network and find that they are highly correlated (the Pearson correlation coefficient of the two sets is 0.9980). For this reason, and because of the more straightforward interpretation, we focus exclusively on $z_p$. The parent dissimilarity values in our data range from 2 to 240.

To test the stability of this measure we also added various amounts of noise to the original weight matrix; for example, adding 1\% noise means that we adjusted each entry by a random number such that its perturbed value is within 1\% of its original value. We then compute the set of shortest-path trees for the perturbed weight matrix, calculate the tree dissimilarities, and then compute the Pearson correlation of the original dissimilarities and the perturbed. The results (0.9995 for 0.1\% noise, 0.9984 for 1\% noise, 0.9937 for 5\% noise) indicate the method is robust against small perturbations, and in addition we do not observe significant changes in the structure of borders determined by the perturbed matrices.

% subsection measuring_tree_distance (end)

\subsection{Hierarchical clustering and borders} % (fold)
\label{sub:hierarchical_clustering}
The measures described above produce a dissimilarity matrix well-suited for use with hierarchical clustering \cite{Everitt:2001p1727}. This technique iteratively groups data points together into clusters that are less and less similar; it begins by identifying the two points with the lowest dissimilarity and grouping them together, then finding the next-most-similar data point or group, and so on. When it is necessary to compare the dissimilarity of one point (or group of points) with another group of points, a linkage function is used. There are several commonly-used linkage functions; we compute single linkages (comparing the shortest distance between two groups), average linkages (the average distance between two groups), and complete linkages (the greatest distance between two groups) and find that the average linkage produces the best fit to our data (Table~\ref{table:comparison}).

\begin{table}
    \begin{center}
    \caption{\label{table:comparison} Cophenetic correlation coefficients~\cite{Sokal:1962p1687} for various linkage functions using parent dissimilarity of trees and inverse weight of links}
    \begin{tabular}{rcc}
        \toprule
        Linkage & $z_p(T_i,T_j)$ & $1/w_{ij}$\\
        \midrule
        single   & 0.6584 & 0.1883\\
        average  & 0.8048 & 0.3757\\
        complete & 0.7197 & 0.1400\\
        \bottomrule
    \end{tabular}
    \end{center}
    
\end{table}

\begin{figure}
    \centering
    \begin{minipage}[c]{0.49\textwidth}
        \includegraphics[width=\textwidth]{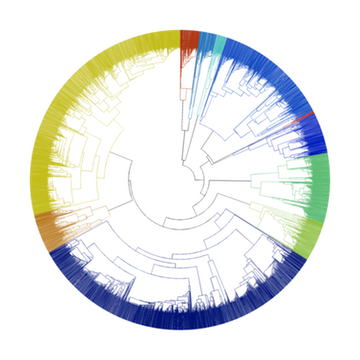}
    \end{minipage} \hfill
    \begin{minipage}[c]{0.4\textwidth}
        \includegraphics[width=\textwidth]{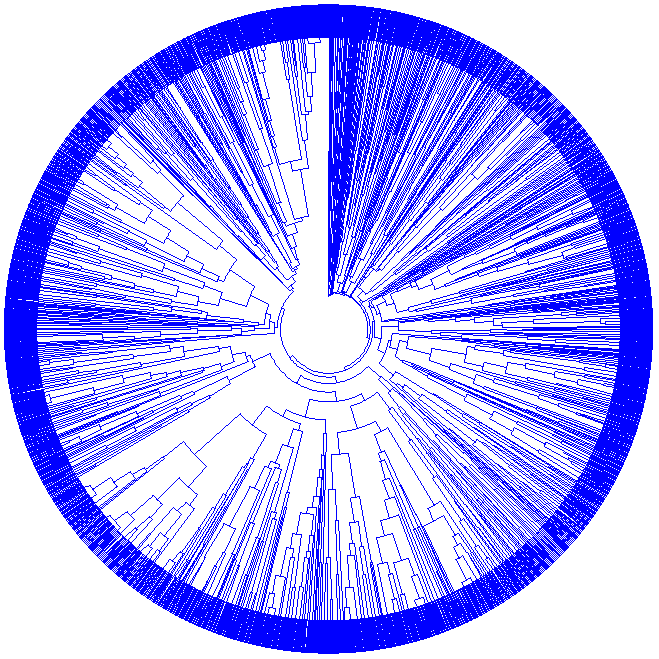}
    \end{minipage}
    \caption{Dendrograms from hierarchical clustering. (\textbf{Left}) Using the parent dissimilarity matrix and average linkage. Colors correspond to a particular community partition depicted in Figure~\ref{fig:dendroMap}. (\textbf{Right}) Using the inverse weight matrix with noise and average linkages. Even inspection by eye reveals immediately that clustering of the inverse weight matrix produces a poor fit to the data, pointing to the need for some type of `pre-conditioning,' here provided by SPT dissimilarity.}
    \label{fig:dendro}
\end{figure}

The result of the hierarchical clustering algorithm is a linkage structure that can be represented graphically with a dendrogram (Figure~\ref{fig:dendro}). The radial lines in the dendrogram represent vertices in our network or groups of vertices, and the arcs represent a link that joins groups together in the hierarchy. The nearer an arc is to the center of the circle, the greater the dissimilarity between the groups joined by the arc.

Each arc corresponds to a geographic border between a set of counties, and the closer the arc is to the center of the circle, the more significant the border. At the outermost level, the dendrogram necessarily puts a border around each individual county, and we threshold at $30\%$ of the height of the tree (corresponding to a dissimilarity $z_p=41.6019$) for the analysis of the WG network.

\begin{figure}
    \centering
        \includegraphics[width=0.75\textwidth]{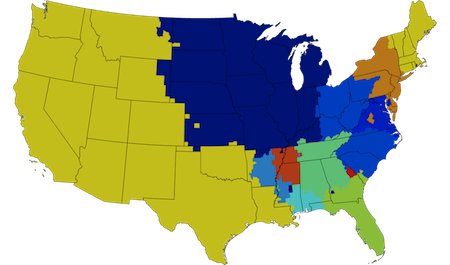}
    \caption{The geographic partition determined by cutting the dendrogram of Figure~\ref{fig:dendro} at a height of 95}
    \label{fig:dendroMap}
\end{figure}

As you can see in Figure~\ref{fig:dendroMap}, the high-level groups identified by this procedure are spatially coherent, but may be divided into spatially disjoint regions at certain heights in the dendrogram.

Hierarchical clustering is also sometimes applied directly to the inverse of the weights, $1/w_{ij}$. We have investigated this method as well and find that it has several shortcomings. First, to apply a hierarchical clustering algorithm requires computing a dissimilarity for every pair of data points; since many pairs of counties are not directly connected by a link in our network ($w_{ij}$ is zero), the inverse does not exist and it is consequently necessary to add some noise to the weight matrix at the very first step, representing pairs of vertices that are `extremely distant' but not disconnected. Second, the linkage structures produced from this approach fit the data poorly (Table~\ref{table:comparison}). Last, one can see by visual inspection of the dendrograms in Figure~\ref{fig:dendro} that this approach does not yield significant information. Comparing the dendrograms for the $z_p$ and $1/w$ matrix, we see that in the shortest-path tree approach most of the links that appear higher in the tree (closer to the center) are linking together two groups that are strongly dissimilar from one another (seen by comparing the height of the parent link to the heights of the children links). In the inverse weight method, this is not true: links high in the tree are linking groups that are quite similar; that is, inverse weight clustering does not identify groups of strongly dissimilar vertices.

\begin{figure}
    \centering
    \includegraphics[type=png,ext=.png,read=.png,width=3in]% 
        {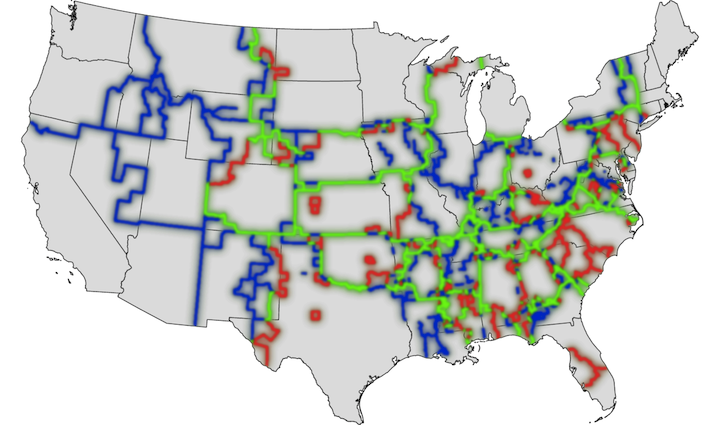}
    \caption{Comparing borders from modularity maximization (blue) with SPT clustering (red) reveals a significant overlap (green). The cumulative topological overlap (see Section~\ref{sec:comparisons}) is $0.5282$ indicating that the SPTD method represents an alternative computational approach to border extraction.}
    \label{fig:mod_spt_overlap}
\end{figure}

Although the method yields a unique sequence of topological segmentations, the observed geographic borders exhibit a strong correlation with those determined by modularity maximization (Figure~\ref{fig:mod_spt_overlap}).

\subsection{Link significance}

\begin{figure}
    \sidecaption
    \includegraphics[width=2in]% 
        {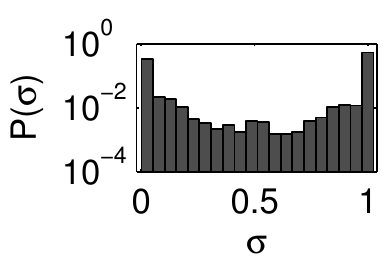}
    \caption{The distribution of link significance $\sigma$, defined for each link as the number of shortest-path trees the link appears in, exhibits a strong bimodal distribution. This implies that SPTD can sort links into important or not, and that $\sigma$ is approximately a binary variable.}
    \label{fig:link_significance}
\end{figure}

The key advantage of this method is that it can systematically extract properties of the network that match the observed borders. A way to demonstrate this is to measure the frequency $\sigma$ at which individual links appear in the ensemble of all SPTs, which is conceptually related to their link betweenness~\cite{Newman:2004p1191}. Computing this \emph{link significance} $\sigma$ for each connection, we find that the distribution $P(\sigma)$ of the network is bimodally peaked (Figure~\ref{fig:link_significance}). This is a promising feature of $P(\sigma)$ as it allows labeling links as either significant or redundant without introducing an arbitrary threshold which is necessary for more continuously distributed link centrality measures. Extracting the group of significant links and constructing a subnetwork from these links only we observe that this subnetwork matches the computed border structure. By virtue of the fact that the most frequently shared links between SPTs are local, short-range connections we see that the SPT boundaries enclose local neighborhoods and that the boundaries fall along lines where SPTs do not share common features. Note that effective metropolitan areas around cities can be detected with greater precision than modularity, although the western US is detected as effectively a single community.

Finally, we performed statistical analyses that quantify the overlap of the effective, mobility-induced borders with those provided by census-related systems. We choose the set of borders separating the states, the borders defined by the districts of the 12~Federal Reserve Banks, and the borders of Economic Areas~\cite{BEA:2010}. We discuss this analysis in more detail in Section~\ref{sec:comparisons}, but briefly, we find a significant correlation with economic boundaries ($p<0.001$, $z$-score $8.024$ for the modularity borders and $p<0.001$, $z$-score $13.29$ for the SPT borders).

% subsection hierarchical_clustering (end)

% section shortest_path_tree_clustering (end)

\section{Significance and comparison of border structures} \label{sec:comparisons}

\subsection{Bootstrapping the Where's George data} \label{sec:boot}

\begin{figure}[hbtp]
	\centering
	\includegraphics[width=\linewidth]{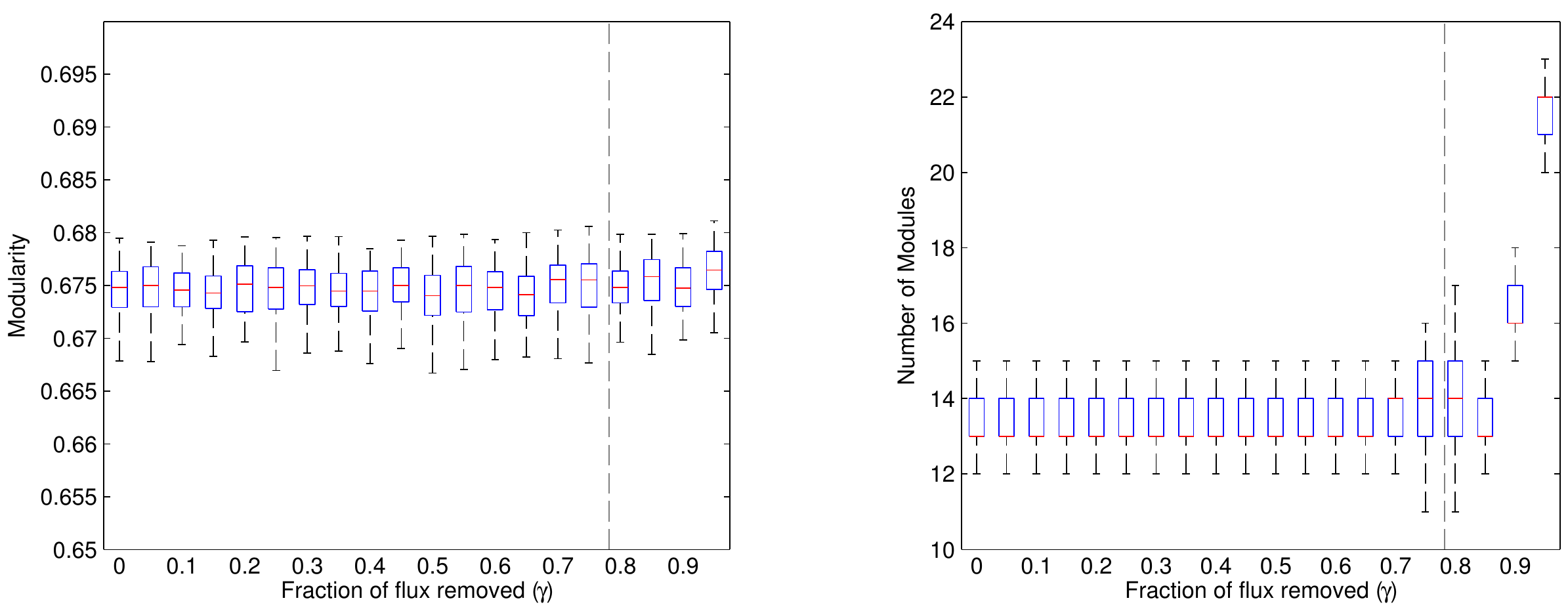}
	\caption{\label{fig:bootSet2}Distributions of modularity values and number of modules for an ensemble of 100~partitions computed for each value of the bootstrapping parameter $\gamma$.  The dashed line corresponds to 78.2\%, the amount of flux ignored if all links shorter than 400\,km would be removed.}
\end{figure}

\begin{figure}[hbtp]
	\mbox{}\hfill$\gamma = 0$\hfill\mbox{}\hfill$\gamma = 0.5$\hfill\mbox{}\\
	\includegraphics[width=.475\linewidth]{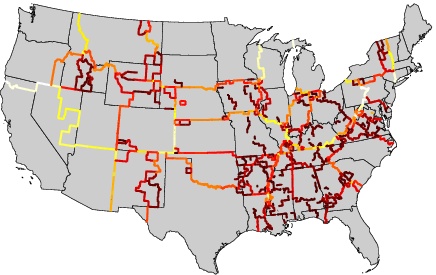}\hfill
	\includegraphics[width=.475\linewidth]{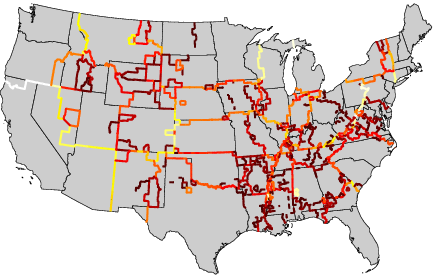}\\[\baselineskip]
	\mbox{}\hfill$\gamma = 0.8$\hfill\mbox{}\hfill$\gamma = 0.95$\hfill\mbox{}\\
	\includegraphics[width=.475\linewidth]{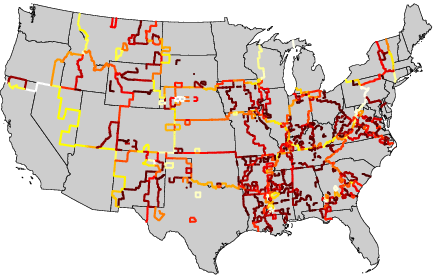}\hfill
	\includegraphics[width=.475\linewidth]{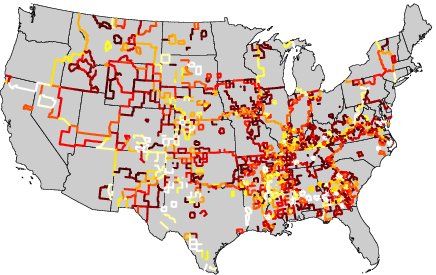}%
	\includegraphics[width=.05\linewidth]{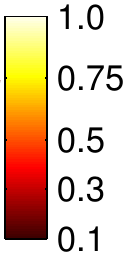}\kern-.05\linewidth
	\caption{\label{fig:bootBounds}Linear superposition of 100~partitions for four different values of the bootstrapping parameter $\gamma$, color-coded according to the fraction of partitions they appear in.}
\end{figure}

In order to test the robustness of our method against random data removal, we performed the following bootstrapping analysis.  Starting with the full dollar bill dataset, and the resulting network weight matrix $W$ with elements $w_{ij}$, we randomly remove single dollar bill reports until the total flux $f = \sum_{i,j} w_{ij}$ is reduced by a factor $\gamma$.  Using this method we constructed several networks for $0\leq\gamma\leq 0.95$ and computed an ensemble of 100~partitions for every value of $\gamma$, using the simulated annealing algorithm described in Section~\ref{sec:modmax}.  We find that the modularity value is unaffected by bootstrapping even if 95\% of the total flux is removed, although the number of modules in each partition rises as the network is thinned out more than 85\% (Figure~\ref{fig:bootSet2}).  Also, the boundary structure emerging from superposition of all partitions is very robust under this procedure (Figure~\ref{fig:bootBounds}).  At 20\% of the original flux ($\gamma = 0.8$) virtually all of the boundaries found in the complete network are still identified, although the sparsity of the data evokes some singular counties.  Even with only 5\% of the flux, when boundaries become more fuzzy, some of the original structures are still detected.

\subsection{Measuring overlap of two boundary networks}

In this section, we describe how to compare boundary networks defined on a planar graph, in our case the county network of the continental US excluding Alaska.

A \emph{boundary network} $b$ is simply given by assigning a nonnegative number $w$ to each edge between adjacent counties: If the two counties are not divided by $b$ then $w=0$. Otherwise $w>0$ implies that the border shared between the two counties has the strength $w$. In Section~\ref{sec:degeneracy} we described how to generate such a boundary network by superposition of many partitions of the Where's George money travel network. We denote this boundary network by the modularity boundaries $b_M$.

We want to quantify how much information the modularity boundaries $b_M$ shares with e.g.~a state network, a random network, or a boundary network generated with another method. 

For this we essentially need to determine the cross-correlation between two boundary networks $b$ and $b'$. However cross-correlation itself is not well-suited for dealing with the non-negativity of the edge weightings, so we calculate a non-centered version of it. The \emph{absolute cross-correlation} of the two boundary networks $b$ and $b'$ is then given by the normalized scalar product of their edge weightings, i.e.~by
$$a(b,b'):=\frac{ (1/|E|) \sum_{e\in E} b(e)b'(e)}
{\sqrt{(1/|E|) \sum_{e\in E} b(e)^2}\sqrt{(1/|E|) \sum_{e\in E} b'(e)^2}}
$$
where $E$ denotes the set of edges connecting adjacent counties. 
This quantity lies between 0 and 1 and equals 1 if and only if the two boundaries are identical up to scaling.

Apart from the upper bound, this quantity however is difficult to judge. In particular, we cannot compare right away two cross-correlations between different networks since $a(\cdot,\cdot)$ might depend on the number of clusters and inhomogeneity of weights etc. We avoid finding a direct interpretation of the absolute cross-correlation by instead considering deviation of observed values against cross-correlations with a null model.

Such null models are used to tell random occurrences of structures from true information. One typically wants to keep some statistics of the network fixed while at the same time randomly sampling from its representational class. This results in the notion of random graphs with certain additional properties such as Erd{\"o}s-R{\'e}nyi~\cite{ErRe59} or Barab{\'a}si-Albert~\cite{BaAl99}.  The key idea is to generate a random network preserving planarity and possible additional information by using the original structure and iteratively changing it by a random local modification. For instance for unweighted networks, a random graph can be generated by `rewiring': two distinct edges and two different vertices contained in either of the two are randomly selected and then swapped. Clearly this operation keeps both degree distributions fixed. After a certain number of iterations, the thus-generated Markov chain produces independent samples of the underlying random graph with given degree distributions~\cite{MaSn02}. This concept has been generalized to weighted graphs~\cite{ZlBiDi09}; in this case it is debatable whether to swap the whole weighted edge or to split up the weight.

\begin{figure}
\includegraphics[width=\columnwidth]{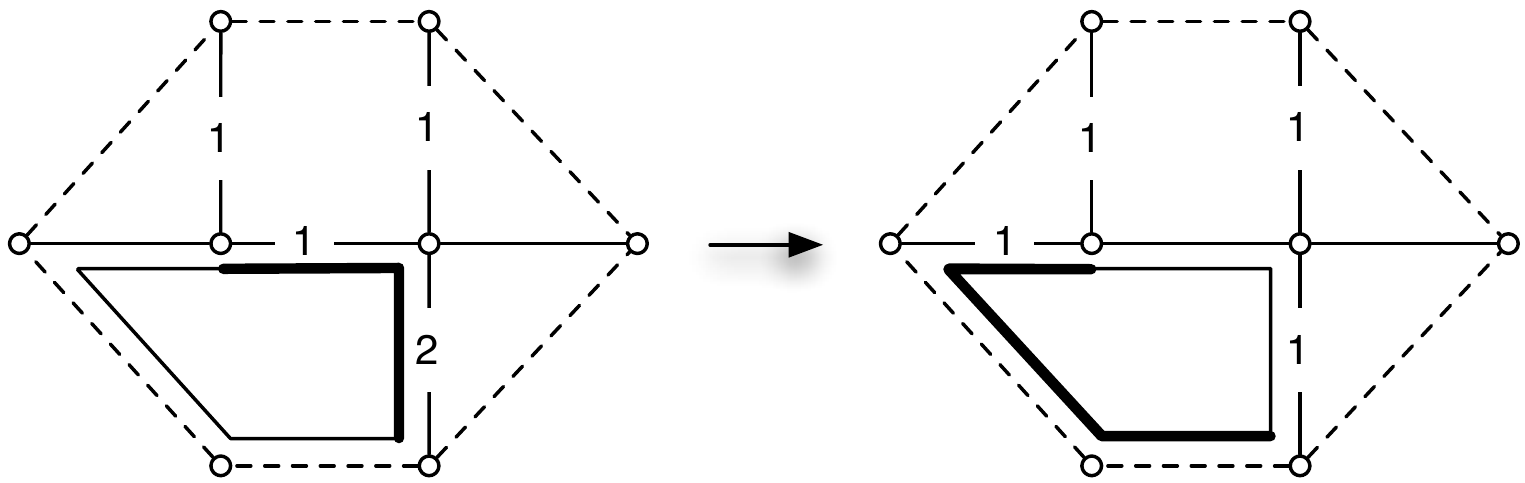}
\caption{\label{fig:planarmove}Local modification of a planar graph. We select the bottom left county to modify. The selected path to modify is shown in bold in the left figure. Its minimal weight is $1$. This is subtracted in the right hand figure, where the complementary path is shown.}
\end{figure}

In our case we search for a randomization of a boundary network i.e.~of a planar, weighted graph. Rewiring as above is not possible since it would destroy planarity. Instead we propose to locally modify the graph at a random county: select a subpath of its boundary and flip it to its complement. In the case of non-trivial weights, we reassign a random number between 0 and the minimal edge weight on the subpath.  We have illustrated this procedure on an example in Figure~\ref{fig:planarmove}.

This procedure is now repeated multiple times until sufficiently de-correlated samples from the original network are produced. In practice, it is common to choose iterations in the range of the number of edges in the network or more. 

\begin{figure}
	\mbox{}\\
	\vspace{-4em}%
	\includegraphics[width=.5\columnwidth]{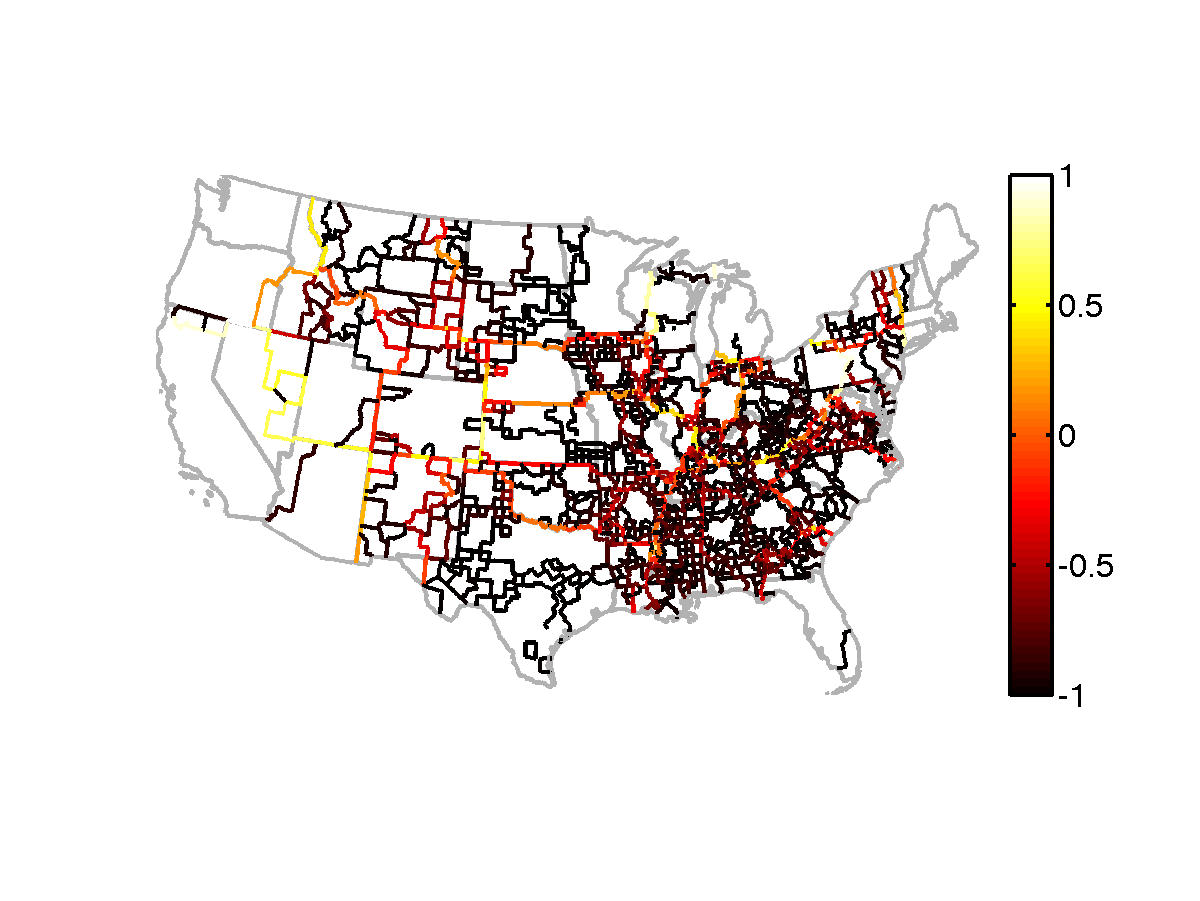}%
	\includegraphics[width=.5\columnwidth]{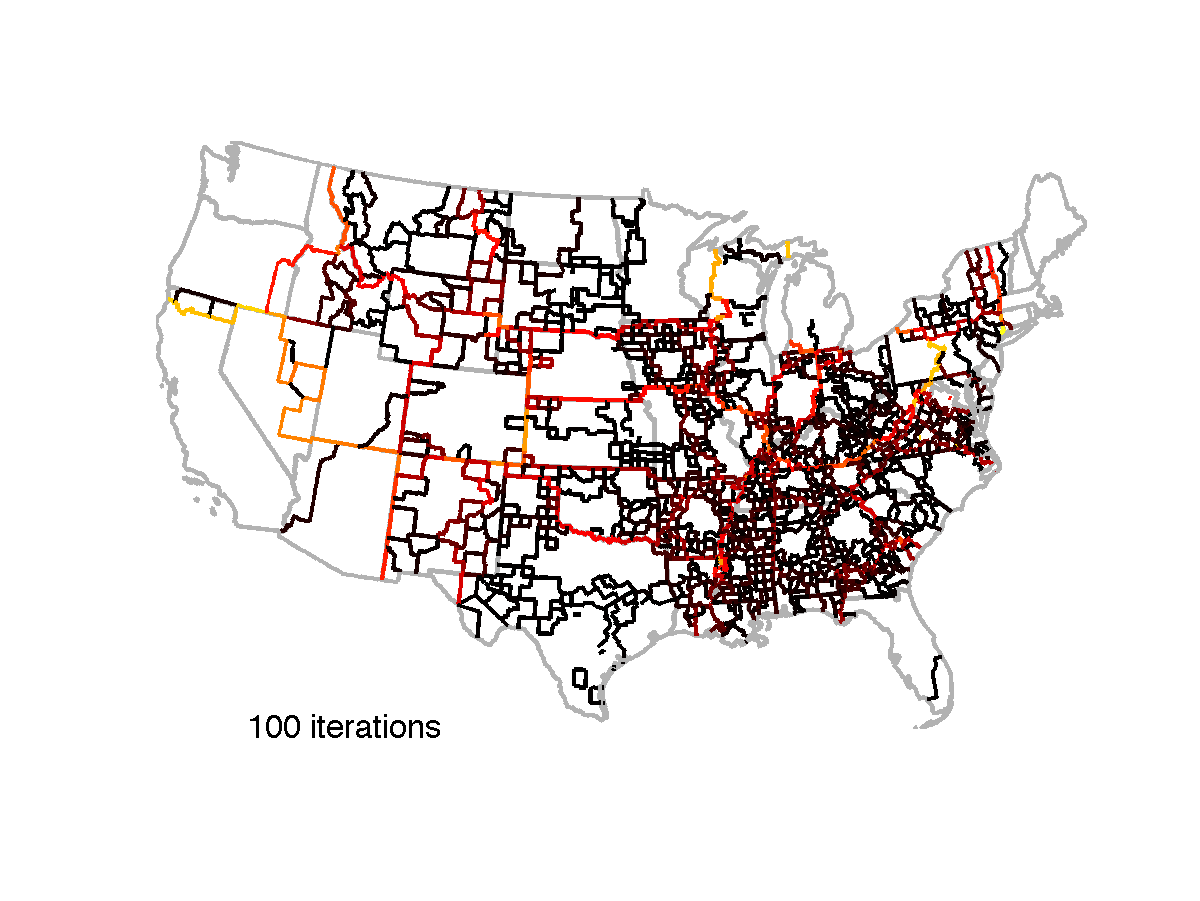}\\
	\vspace{-4em}%
	\includegraphics[width=.5\columnwidth]{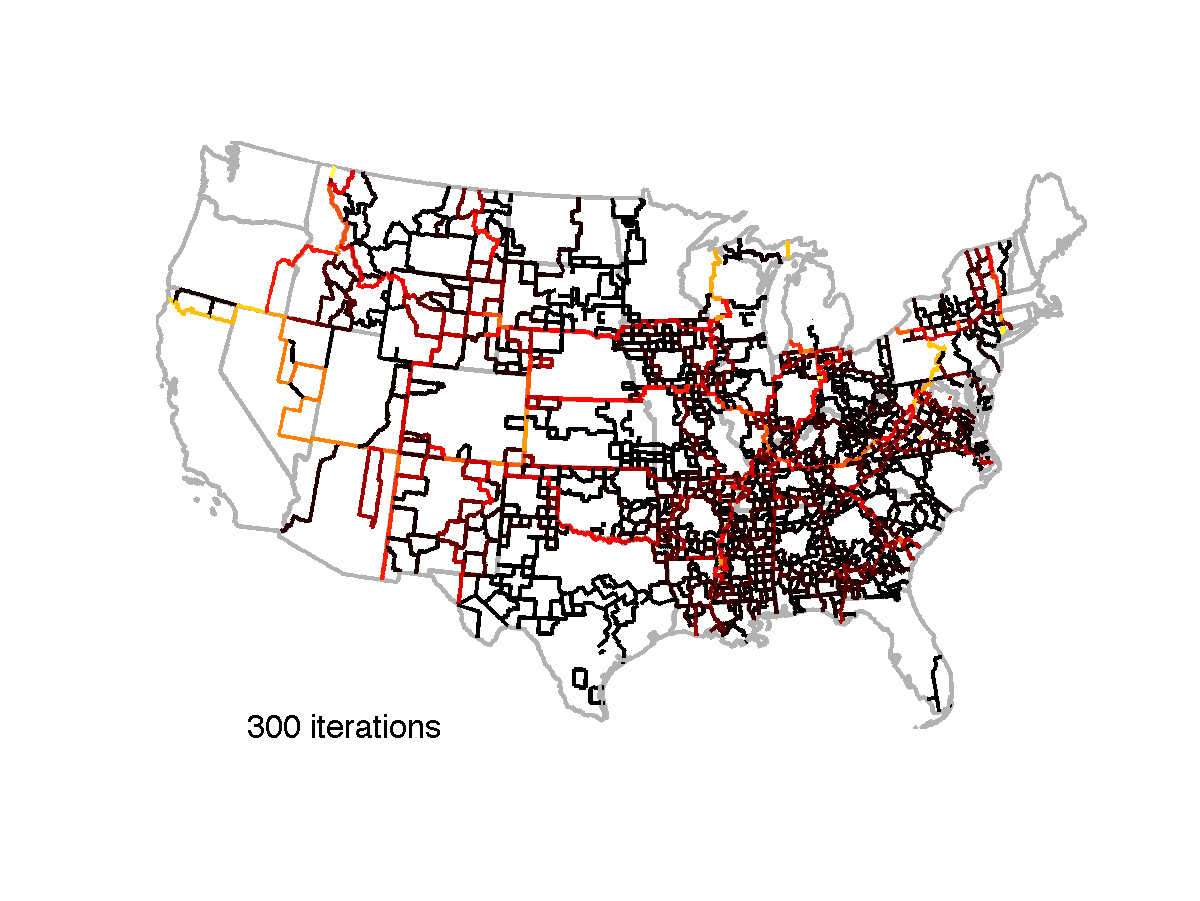}%
	\includegraphics[width=.5\columnwidth]{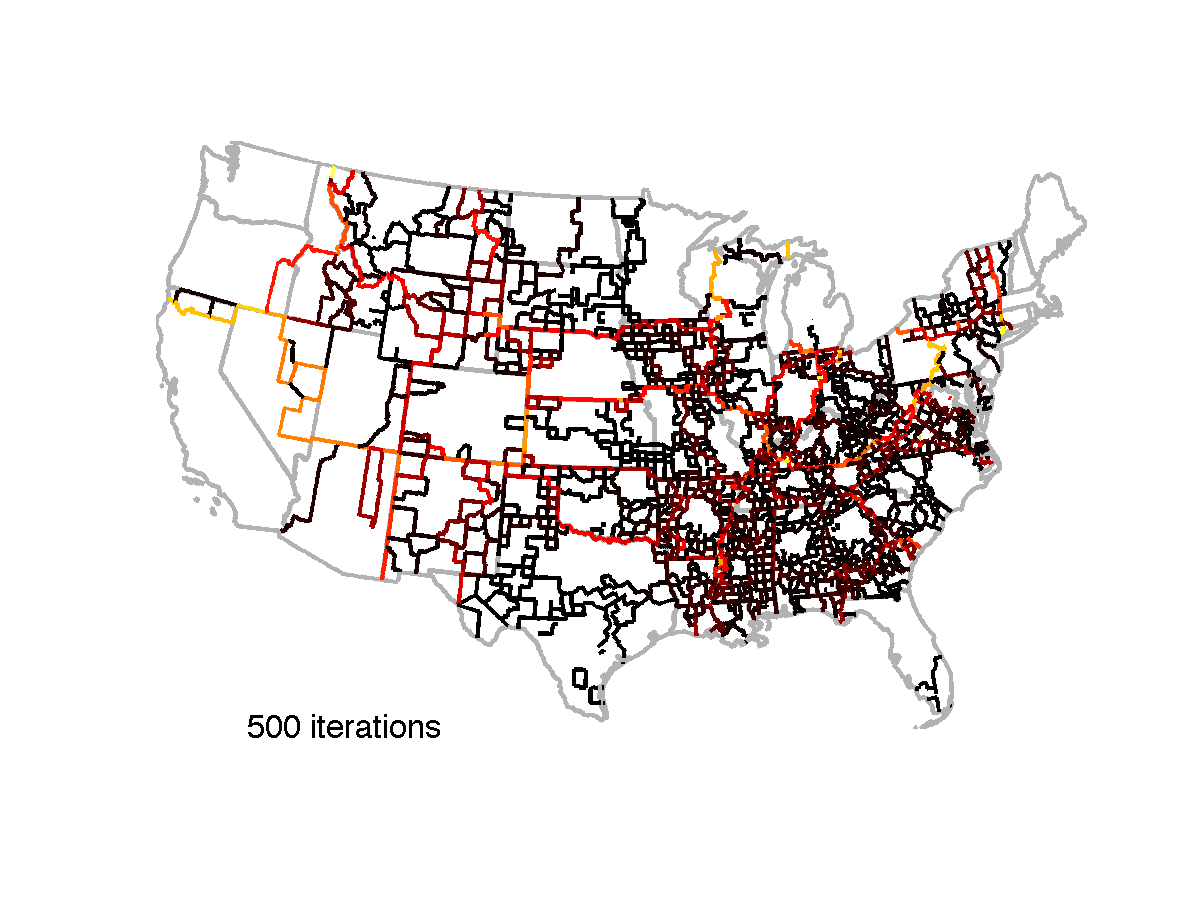}\\
	\vspace{-4em}%
	\includegraphics[width=.5\columnwidth]{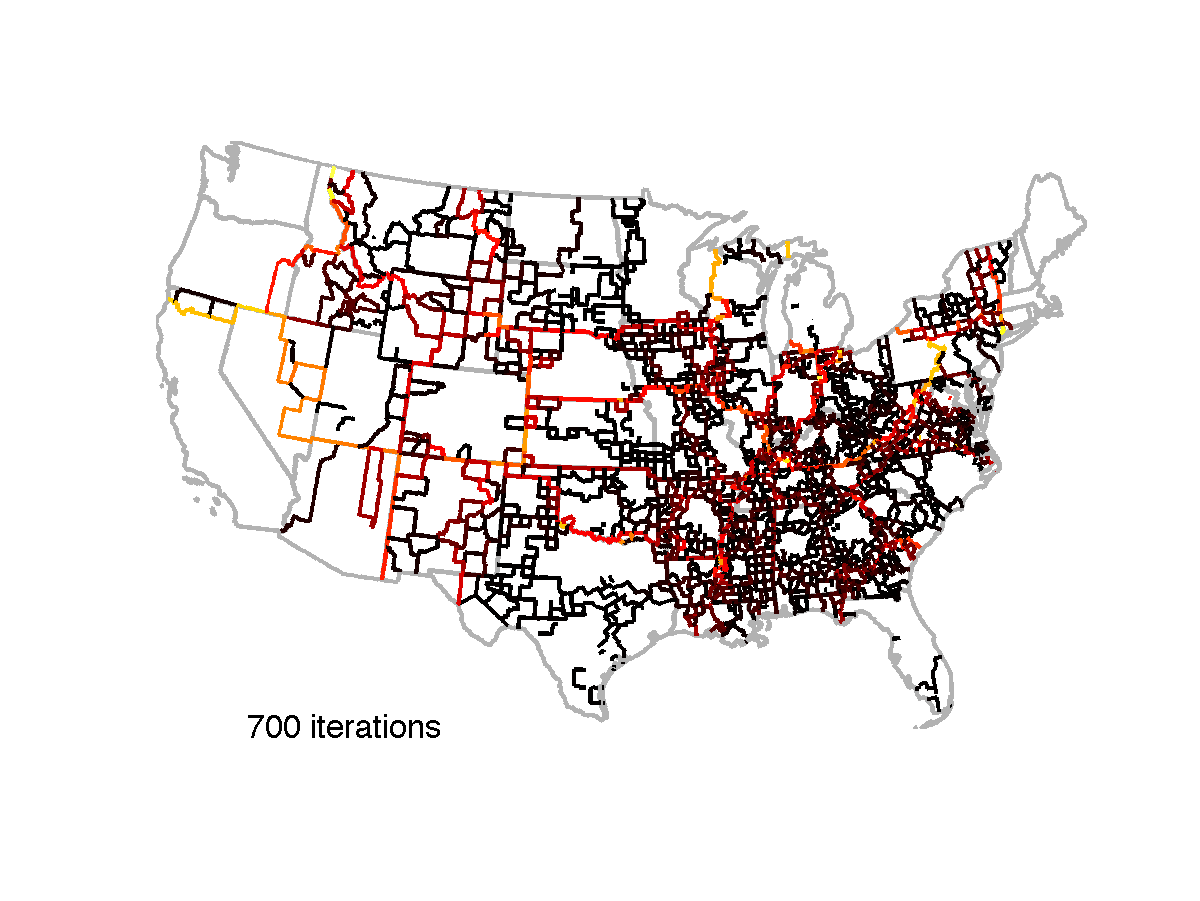}%
	\includegraphics[width=.5\columnwidth]{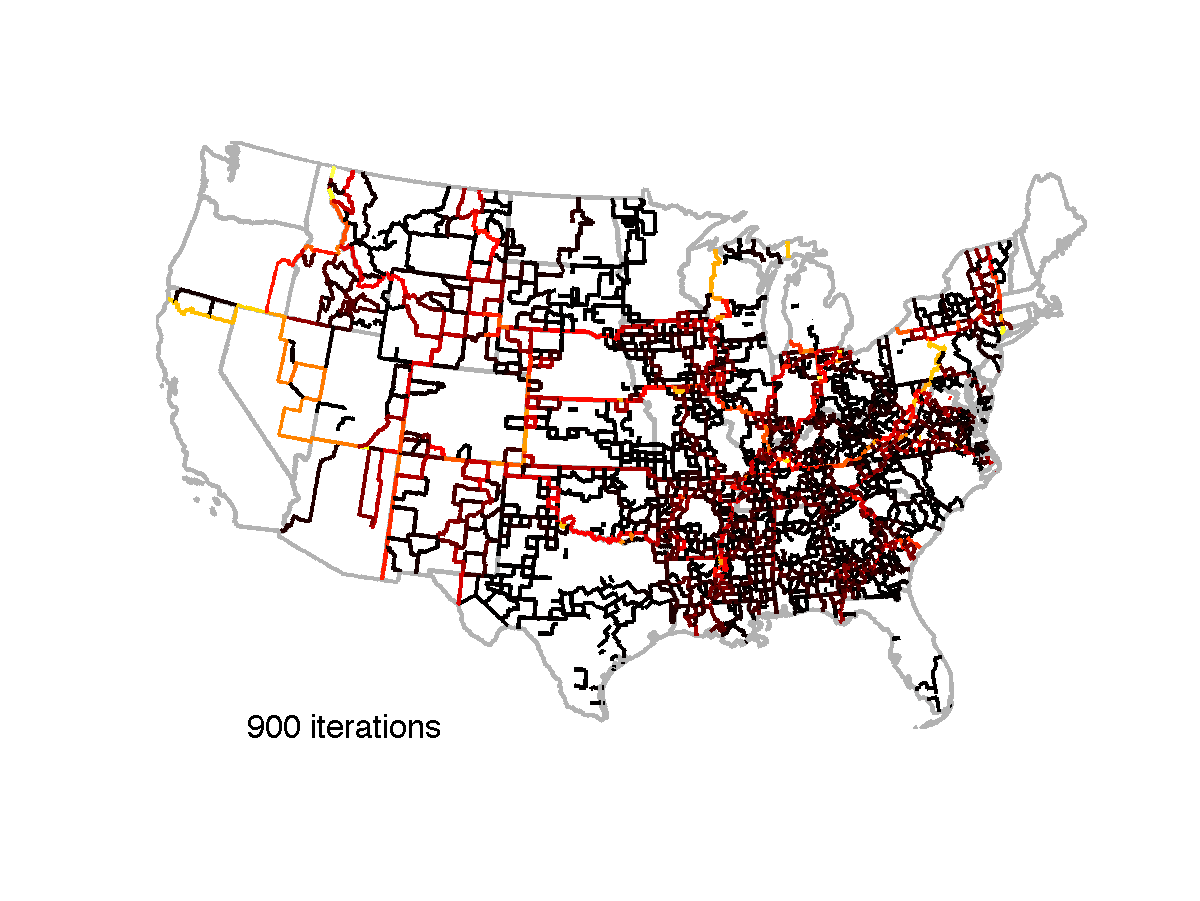}\\
% \includegraphics[width=.5\columnwidth]{images-randomization/boundaries-randomization-it0}%
% \includegraphics[width=.5\columnwidth]{images-randomization/boundaries-randomization-it100}\\
% \vspace{-4em}%
% \includegraphics[width=.5\columnwidth]{images-randomization/boundaries-randomization-it200}%
% \includegraphics[width=.5\columnwidth]{images-randomization/boundaries-randomization-it300}\\
% \vspace{-4em}%
% \includegraphics[width=.5\columnwidth]{images-randomization/boundaries-randomization-it400}%
% \includegraphics[width=.5\columnwidth]{images-randomization/boundaries-randomization-it500}\\
% \vspace{-4em}%
% \includegraphics[width=.5\columnwidth]{images-randomization/boundaries-randomization-it600}%
% \includegraphics[width=.5\columnwidth]{images-randomization/boundaries-randomization-it700}\\
% \vspace{-4em}%
% \includegraphics[width=.5\columnwidth]{images-randomization/boundaries-randomization-it800}%
% \includegraphics[width=.5\columnwidth]{images-randomization/boundaries-randomization-it900}\\
\caption{\label{fig:WGrand1000}Randomization of the modularity boundary network $b_M$. The original network and the first 900 iterations are shown.}
\end{figure}

\begin{figure}\centering
\includegraphics[width=.5\columnwidth]%
    {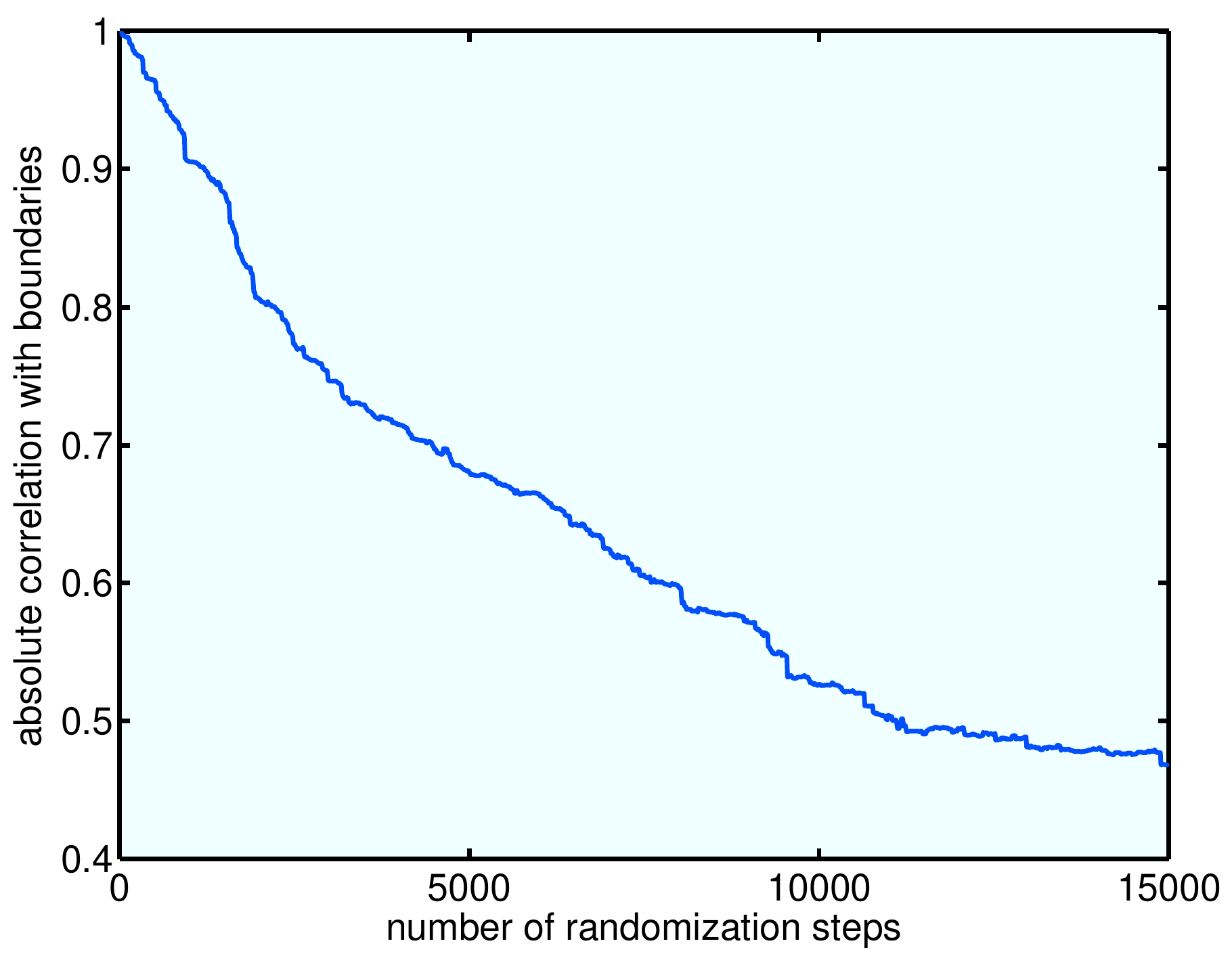}
\caption{\label{fig:WGabscor}Absolute cross-correlation of the randomized network after the given number of iterations with the modularity boundary network $b_M$.}
\end{figure}

\begin{figure}\centering
    \includegraphics[width=.8\columnwidth]%
        {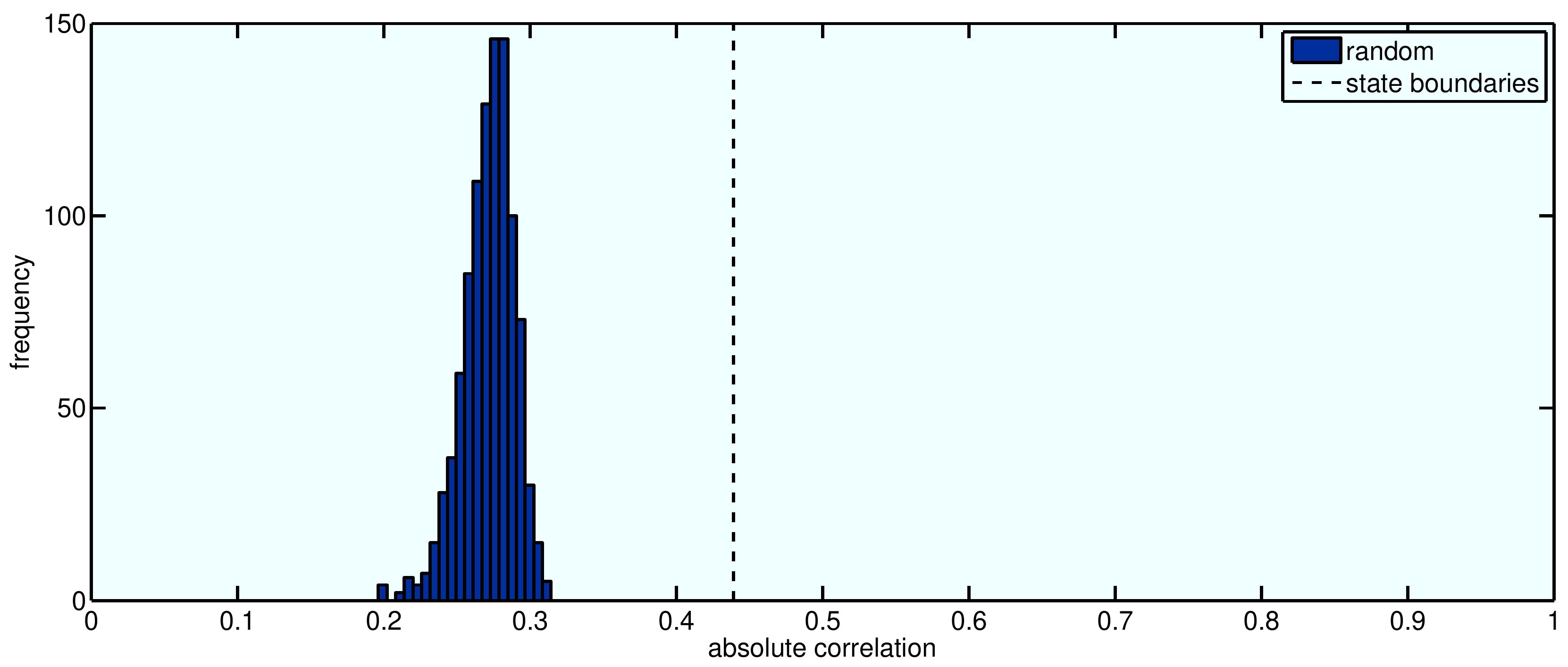}\\
    \textbf{a} absolute correlations of the null model and $b_M$ with the state boundaries\\[\baselineskip]
    \includegraphics[width=.8\columnwidth]%
        {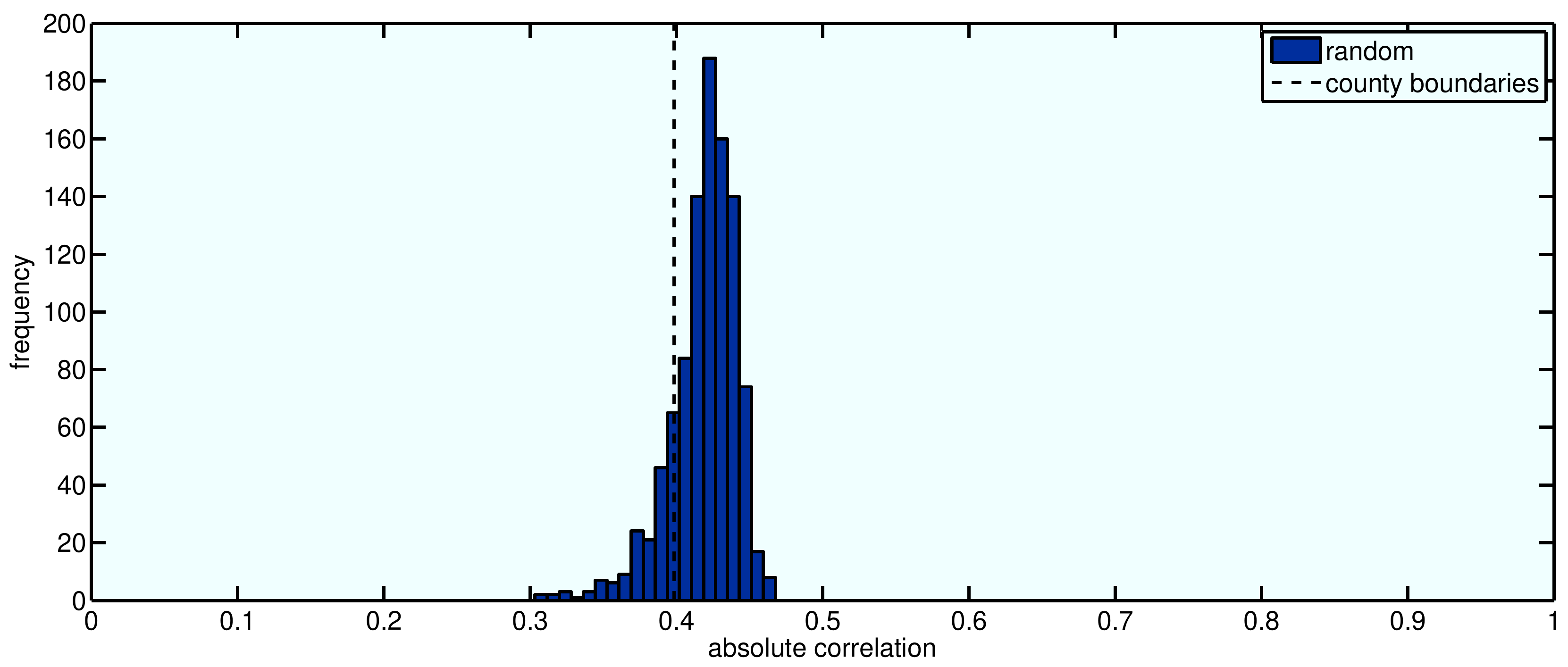}\\
    \textbf{b} absolute correlations of the null model and $b_M$ with the county boundaries
    \caption{\label{fig:WGfeatures}Absolute cross-correlation of state and county boundaries when compared with a null model based on the modularity boundary network $b_M$.}
\end{figure}

\subsection{Randomization of the mean partition boundary of the Where's George network}

In order to test for significances of calculated similarities, we build a random model of the mean partition boundary by generating $1000$ random networks using the above algorithm with $>15000$ successful iterations for each random network. The corresponding maps for the first $900$ iterations are shown in Figure~\ref{fig:WGrand1000}. Clearly the original structure in the boundary network is increasingly diluted, and after $>10000$ iterations becomes stably random.

This can be seen by calculating the absolute cross-correlation $a(b_M,b_R)$ of the modularity boundary network $b_M$ with the random networks $b_R$, when increasing the number of iterations, see Figure~\ref{fig:WGabscor}. We observe convergence to roughly $0.5$ after about 10000 steps. This lies well in the range of random correlation with a mean of $0.49$ and a standard deviation of $0.028$, see histogram in Figure~\ref{fig:WGfeatures}(a). This implies that the randomization procedure converges to a set of random boundary networks, which can now be used to put calculated autocorrelations into perspective against this null model.

\subsection{Significances when comparing boundary networks with the null model}

We describe and quantify overlap of the estimated modularity boundaries $b_M$ with other political or social boundaries. As described before, we can quantify overlap by determining the absolute cross-correlation $a(b,b_M)$. In order to determine interpretable numbers, we compare this value to correlations with random boundaries $b_R$ from a null model.

We now determine significance of coincidence of the modularity boundary network $b_M$ and the SPT boundary network $b_S$ with 
\begin{itemize}
\item modularity boundaries $b_M$,
\item state boundaries,
\item county boundaries (to test for sensitivity of the method against number of communities),
\item boundaries resulting from the SPT algorithm $b_S$,
\item boundaries determined on the gravity model,
\item boundaries determined on long-range distances only,
\item federal reserve district boundaries (FRB), and
\item economic area boundaries (\url{http://www.bea.gov}).
\end{itemize}
The significance is calculated by replacing $b_M$ and $b_S$, respectively, by elements from the corresponding null model.

For illustration we show two histograms and actual values for state and county boundaries in Figure~\ref{fig:WGfeatures}. Clearly the random cross-correlations are quite different, which means that we have to interpret the actual values of $0.439$ and $0.398$ differently as well. Indeed it turns out that the state value is far from the mean random cross-correlation $0.272\pm0.018$, whereas the county one is not ($0.419\pm0.023$). Indeed, the empirical p-values, determined as the fraction of random correlations above the observed true one, is 0 in the former and 0.84 in the latter case.

In order to compare cases with large deviation from the distribution, we determine the $z$-score i.e. the distance of the absolute cross-correlation from the mean of the null model normalized by the standard deviation:
$$z(b):=\frac{a(b,b_M)-E(a(b,b_R))}{\std(a(b,b_R))},$$
where $E$ denotes mean and $\std$ standard deviation. In the state case, this z-score is very high, 9.46, which means that the observed correlation is more than 9 standard deviations away from the random mean. In contrast the county z-score is 0.90, which means that the observation is within one standard deviation and hence not significant.

We summarize the calculated cross-correlations in Tables~\ref{table:compWG} and~\ref{table:compSPT} for $b_M$ and $b_S$. 

\begin{table}
\renewcommand{\arraystretch}{1.3}
\caption{Comparing boundary overlaps for various boundary networks with the modularity boundaries $b_M$ and the corresponding null model $b_R$ using absolute cross-correlation $a$.}
\label{table:compWG}
\begin{center}
\begin{tabular}{rcccc}
\toprule
boundary network & $a(\cdot,b_M)$ & $a(\cdot,b_R)$ & p-value & z-score\\
\midrule
modularity boundaries	& 1.000	 & $0.495\pm0.028$ & $<10^{-3}$ & 18.15\\
%quick communities & 0.783 & $0.429\pm0.025$ & $<10^{-3}$ &  14.01\\
SPT communities & 0.552 & $0.385\pm0.024$ & $<10^{-3}$ & 7.03\\
state boundaries	 & 0.439	 & $0.272\pm0.018$	& $<10^{-3}$ & 9.46\\
county boundaries & 0.398 & $0.419\pm0.023$ &	0.84	 & 0.90 \\
%OLD grav model: gravity boundaries & 0.266 & $0.262\pm0.020$ &	0.44	 & 0.20 \\
gravity boundaries & 0.260 & $0.253\pm0.019$ &	0.35	 & 0.40 \\
large-range network boundaries & 0.198 & $0.181\pm0.017$ &	0.14	 & 1.02\\
federal reserve district boundaries	& 0.377 & $0.227\pm0.019$ & $<10^{-3}$ & 7.91\\
economic area boundaries &0.452 & $0.307\pm0.018$ & $<10^{-3}$ & 8.024 \\
\bottomrule
\end{tabular}
\end{center}
\end{table}

\begin{table}
\renewcommand{\arraystretch}{1.3}
\caption{Comparing boundary overlaps for various boundary networks with the SPT-based boundary $b_S$ and the corresponding null model $b_R$ using absolute cross-correlation $a$.}
\label{table:compSPT}
\begin{center}
\begin{tabular}{rcccc}
\toprule
boundary network & $a(\cdot,b_S)$ & $a(\cdot,b_R)$ & p-value & z-score\\
\midrule
modularity boundaries	& 0.552	 & $0.251\pm0.013$ & $<10^{-3}$ & 22.55\\
%quick communities & 0.539 & $0.244\pm0.014$ & $<10^{-3}$ &  21.44\\
SPT communities & 1.000 & $0.367\pm0.0164$ & $<10^{-3}$ & 40.63\\
state boundaries	 & 0.358	 & $0.220\pm0.0138$	& $<10^{-3}$ & 10.99\\
county boundaries & 0.569 & $0.562\pm0.016$ &	0.36 & 0.44 \\
%OLD grav model: gravity boundaries & 0.305 & $0.275\pm0.016$ &	0.03	 & 1.81 \\
gravity boundaries & 0.305 & $0.260\pm0.016$ &	0.002 & 2.73 \\
large-range network boundaries & 0.257 & $0.199\pm0.015$ &	$<10^{-3}$ 	 & 3.94\\
federal reserve district boundaries	& 0.307 & $0.159\pm0.013$ & $<10^{-3}$ & 11.79\\
economic area boundaries & 0.492 & $0.318\pm0.013$ & $<10^{-3}$ & 13.29 \\
\bottomrule
\end{tabular}
\end{center}
\end{table}

\subsection{Discussion}

For the state and the SPT boundaries we observe a strong deviation from the null model when comparing against the modularity boundaries. So we can conclude that both state boundaries and SPT boundaries are more similar to $b_M$ than expected by chance with a p-value $<10^{-3}$.

This is not the case for the gravity model, the county boundaries and the long-range model. In these cases, the cross-correlation with $b_M$ is not larger than with a random model (p-value $\approx 0.44$, $\approx 0.84$ and $\approx 0.14$). This means that they do not significantly coincide with $b_M$.

The absolute cross-correlation of the FRB boundaries with $b_M$ is $a(b_F,b_M)=0.38$, which is significantly high when compared with the null model, which exhibits cross-correlations of only $a(b_F,b_R)=0.23\pm0.019$.  We observe a strong deviation from the null model and can therefore conclude that the FRB boundaries are more similar to $b_M$ than expected by chance with a p-value $<10^{-3}$. 

The corresponding $z$-score equals $7.91$, which is lower than the one for states (9.46). This implies that the modularity boundaries' overlap with the states is larger than the one with the FRB boundaries.

We interpret the results on the FRB boundaries when compared with $b_M$ as follows:
\begin{itemize}
\item 
The structure of $b_M$ may be (partially) due to political structure i.e.~result from $b_S$ or due to additional money transport within FRB districts i.e.~correlate with $b_F$. Since both $b_S$ and $b_F$ share strong similarities, in each of the two situations, we would see overlap with both boundaries, so we can only judge strength of overlap with respect to the other boundary. 
\item
We quantified strength of overlap by deviation from the null model, and the corresponding z-score was more than 1.5 standard deviations higher for the state model. This stronger overlap of states with $b_F$ therefore favors the first hypothesis i.e.~the situation that political boundaries are a stronger factor for the pattern observed in $b_M$. In the case of dominance of the second hypothesis, we would instead expect to still see overlap with state boundaries, but less overlap than with the FRB ones.
\end{itemize}

\bibliographystyle{spmpsci}
\bibliography{boundaries}

\end{document}